\documentclass[aps,prd,reprint,preprintnumbers,superscriptaddress,showpacs,twocolumn]{revtex4-1}
\usepackage{latexsym,graphicx,amssymb,amsmath,mathrsfs}
\usepackage{setspace,bm}
\usepackage[breaklinks, colorlinks=true, pdfstartview=FitV, linkcolor=red, citecolor=blue, urlcolor=blue]{hyperref}
\usepackage[usenames]{color}
\usepackage{latexsym}
\usepackage{epstopdf}
\usepackage{mathtools}
\usepackage{amssymb}

\newcommand{\Slash}[1]{{\ooalign{\hfil/\hfil\crcr$#1$}}}

\allowdisplaybreaks[1]
\usepackage[normalem]{ulem}  % \sout{old text} for strikeout
\renewcommand\sout{\bgroup \color{red} \ULdepth=-.5ex \ULset}

\begin{document}
%\preprint{****-**-**}

%\title{Roles of Roberge-Weiss periodicity for
%confinement-deconfinement transition}
\title{
Roberge-Weiss periodicity, canonical sector and modified Polyakov-loop
}

\author{Kouji Kashiwa}
\email[]{kashiwa@fit.ac.jp}
\affiliation{Fukuoka Institute of Technology, Wajiro, Fukuoka 811-0295,
Japan}

\author{Hiroaki Kouno}
\email[]{kounoh@cc.saga-u.ac.jp}
\affiliation{Department of Physics, Saga University, Saga 840-8502,
Japan}

\begin{abstract}
 To obtain deeper understanding of QCD properties at finite temperature,
 we consider the Fourier decomposition of the grand-canonical partition
 function based on the canonical ensemble method via the imaginary chemical
 potential.
 Expectation values are, then, represented by summation over each canonical
 sector.
 We point out that the modified Polyakov-loop can play an important role
 in the canonical ensemble;
 for example, the Polyakov-loop paradox which is known in the
 canonical ensemble method can be evaded by considering the quantity.
 In addition, based on the periodicity issue of the
 modified Polyakov-loop at finite imaginary chemical potential, we can
 construct the systematic way to compute the dual quark condensate which
 has strong unclearness in its foundation in the presence of dynamical
 quarks so far.
\end{abstract}

\maketitle

\section{Introduction}

Extensive studies for exploring the phase diagram of quantum
chromodynamics (QCD) have been done at finite temperature ($T$) and
real chemical potential ($\mu_\mathrm{R}$).
In the study of QCD, the lattice QCD simulation is a
powerful and gauge invariant approach to investigate its
non-perturbative nature.
Lattice QCD simulation, however, has the well-known sign problem at
finite $\mu_\mathrm{R}$;
see Ref.~\cite{deForcrand:2010ys} as an example.
Therefore, several approaches have been proposed so far such as
the Taylor expansion
method~\cite{Allton:2005gk,*Gavai:2008zr},
the reweighting
method~\cite{Fodor:2001au,*Fodor:2001pe,*Fodor:2004nz,*Fodor:2002km},
the analytic continuation
method~\cite{deForcrand:2002hgr,*deForcrand:2003vyj,*DElia:2002tig,*DElia:2004ani,*Chen:2004tb},
the canonical ensemble
method~\cite{Hasenfratz:1991ax,*Alexandru:2005ix,Kratochvila:2006jx,*deForcrand:2006ec,*Li:2010qf}
and so on.
However, the application range of these approaches are still limited in
the small $\mu_\mathrm{R}/T$ region.

In comparison, the imaginary chemical potential
($\mu_\mathrm{I}$) region does not have the sign problem and thus
we can perform the lattice QCD simulation, exactly.
Interestingly, the $\mu_\mathrm{I}$ region
can play a crucial role to understand QCD properties at finite
$\mu_\mathrm{R}$ via the analytic
continuation method and the canonical ensemble method.
Thus, this region is very useful.
In addition, it has been recently proposed that the structure of QCD at finite
$\mu_\mathrm{I}$ may be related with the confinement-deconfinement
transition~\cite{Kashiwa:2015tna,*Kashiwa:2016vrl,*Kashiwa:2017yvy} based
on the analogy of the topological order at
$T=0$~\cite{Wen:1989iv,Sato:2007xc}.
Because of these reasons, it is interesting and important to know detailed
properties of QCD at finite $\mu_\mathrm{I}$.
%In this study, we concentrate on the investigation based on the
%canonical ensemble method to understand detailed properties of QCD at
%finite $\mu_\mathrm{I}$.

The canonical ensemble method which is strongly related with the imaginary
chemical potential region is the interesting and well known
method: there is the mathematical reason that
the canonical partition functions can be constructed from the
grand-canonical partition function with $\mu_\mathrm{I}$
via the Fourier transformation and the fugacity
expansion~\cite{Roberge:1986mm}.
This approach has been applied to the lattice QCD simulation and
successfully performed in moderate and high temperature regions.
The canonical ensemble method, however, has the Polyakov-loop
paradox~\cite{Kratochvila:2006jx}.
This paradox induces the problem that the Polyakov loop cannot be
used as the order parameter/indicator in the canonical ensemble method
because the Polyakov-loop becomes  always zero at any $T$.

It is easily to understand that
the Polyakov-loop paradox is induced from the fact that the
Polyakov loop does not have the Roberge-Weiss (RW) periodicity;
for example see Fig.~\ref{Fig:F-q}.
In other words, the Polyakov loop is no longer good quantity to
characterize the imaginary chemical potential region.
Thus, we employ the {\it modified Polyakov-loop} in this
article; it is known as the RW periodic quantity and can describe
the RW transition nature.
We show that the Polyakov-loop paradox can be evaded by using the
modified Polyakov-loop in this study;
we can reproduce the Polyakov loop via the canonical ensemble
method.
In addition to resolve the Polyakov-loop paradox, we can clarify the
foundation of the dual quark condensate~\cite{Bilgici:2008qy}
from the RW periodicity issue of the modified
Polyakov-loop;
this quantity has been proposed as the order parameter or the indicator
of the deconfinement transition, but it has strong unclearness in its
determination so far.

This paper is organized as follows.
In the next section, we summarize important properties of QCD at finite
imaginary chemical potential.
Section \ref{Sec:PNJL} shows the possible QCD effective model which
reproduces the QCD properties at finite imaginary chemical potential;
we need the model to demonstrate our observation.
The Polyakov-loop paradox in the canonical
ensemble method is explained in Sec.~\ref{Sec:ce}.
In Sec.~\ref{Sec:mpl}, we discuss how the Polyakov-loop paradox can be
evaded by using the modified Polyakov-loop.
We show the theoretical foundation of the dual quark condensate from
the RW periodicity of the modified Polyakov-loop in
Sec.~\ref{Sec:dqc}.
Section~\ref{Sec:Summary} is devoted to summary.

\section{Structure of QCD at finite imaginary chemical potential}
\label{Sec:IC}

The QCD grand-canonical partition function (${\cal Z}_\mathrm{GC}$)
has following properties at finite $\theta := \mu_\mathrm{I}/T$;
\begin{description}
  \item[1. The Roberge-Weiss (RW) periodicity]
        The imaginary chemical potential can be transformed into the quark
        temporal boundary condition and then the $\mathbb{Z}_3$-transformation
        which induces the $\mathbb{Z}_3$ factor to the quark field can be
        also absorbed into the boundary condition.
        Consequently, we have
	\begin{align}
	 {\cal Z}_\mathrm{GC}(T,\theta)
	 = {\cal Z}_\mathrm{GC}
	   \Bigl(T,\theta+\frac{2\pi k}{N_\mathrm{C}} \Bigr),
	\end{align}
	where $N_\mathrm{c}$ is the number of color and $k \in \mathbb{N}$.
        This periodicity of ${\cal Z}_\mathrm{GC}$ causes the
        $2\pi/N_\mathrm{c}$ periodicity of several thermodynamic
        quantities such as the pressure, the entropy and the quark number density.
        This special $2\pi/N_\mathrm{c}$ periodicity is so called the
	{\it RW periodicity}~\cite{Roberge:1986mm}.

  \item[2. Existence of $\mathbb{Z}_{N_\mathrm{c}}$ images]
        It is well known that there are $\mathbb{Z}_{N_\mathrm{c}}$
	images at high $T$. These images are corresponding to minimum of
	the thermodynamic potential and one of them becomes the global
        minima in certain range of $\theta$.
	However, different minima which is the local minima of
        the thermodynamic potential in the certain range of
	$\theta$ can become the global minima
        when the $\theta$ range is changed.
	The images can be characterized by the phase of the Polyakov
        loop ($\phi$):
	For example,
	\begin{description}
	 \item[Non-trivial] ~~$\phi=2\pi/3$ for $\theta=[-\pi,-\pi/3]$
	 \item[Trivial] \hspace{0.8cm}~~$\phi=0$ for $\theta=[-\pi/3,\pi/3]$
	 \item[Non-trivial] ~~$\phi=-2\pi/3$ for $\theta=[\pi/3,\pi]$
	\end{description}
	are all possible $\mathbb{Z}_{N_\mathrm{c}}$ images for
	$N_\mathrm{c}=3$ at sufficiently high $T$.
	The realistic system exists with the trivial
	$\mathbb{Z}_{N_\mathrm{c}}$ image where
	$\Phi \in \mathbb{R}$.
        These images, however, vanished at low $T$ and then only
	one minimum appears, but the RW
	periodicity still exists.

  \item[3. The RW transition]
        The origin of the RW periodicity at low and high $T$ are
        perfectly different.
        At low $T$, hadronic contributions are quite strong and then the
        RW periodicity is induced by the baryonic fugacity,
        $e^{\pm 3i\theta}$, and then there is no singularity along $\theta$
        direction.
        On the other hand, the quark fugacity, $e^{\pm i(g A_4/T+\theta)}$,
        jumps into the game at high $T$ where $g$ is the gauge coupling
        constant.
        In this case, the RW periodicity is induced via the
        $\mathbb{Z}_3$ images;
        i.e. the global minima of the thermodynamic potential switches to
        another minimum when we across $\theta=(2k-1)\pi/3$ and
	then the singularities arise in thermodynamic quantities.
        This singularity induces the phase transition along $\theta$
        direction and it is so called the {\it RW
	transition}~\cite{Roberge:1986mm}.

 \item [4. RW endpoint] The RW endpoint is the endpoint of the
	first-order RW transition.
	Since the singularities appear when the RW transition happen,
	there should be the endpoint at certain $T$ which is usually
	denoted by $T_\mathrm{RW}$.
	The order of the endpoint may depend on the number of flavor
	with physical quark mass;
	some lattice QCD simulations indicate that the order is
	first-order and then the RW endpoint is the
	triple-point in the two- and three-flavor systems
	\cite{deForcrand:2010he,D'Elia:2009qz,Bonati:2010gi}.
        However, some lattice QCD simulations
	indicate the second-order RW endpoint even below
	the physical quark
	mass in the $2+1$ flavor
	system~\cite{Bonati:2016pwz,Bonati:2018fvg,Goswami:2018qhc}.
\end{description}
Based on these properties, several studies for the
confinement-deconfinement transition have been done;
the first attempt is Ref.~\cite{Weiss:1987mp}, some related discussions
are show in
Refs.~\cite{Kashiwa:2015tna,*Kashiwa:2016vrl,*Kashiwa:2017yvy} and this
paper.
Also, some of them have been used to investigate
the correlation between the confinement-deconfinement transition and the
chiral phase transition via the state-of-art anomaly
matching~\cite{kikuchi2018t,Yonekura,Nishimura:2019umw}.

\section{Possible QCD effective model}
\label{Sec:PNJL}

To demonstrate our observation as shown later,
we need the effective model of QCD which reproduces QCD
properties at finite $\mu_\mathrm{I}$.
The Polyakov-loop extended Nambu--Jona-Lasinio (PNJL)
model~\cite{Fukushima:2003fw} is the famous
and promising effective model for this purpose.
The PNJL model can approximately describe the deconfinement and chiral
phase transitions at the same time by introducing non-perturbative
effects through the NJL model and Polyakov-loop effective
potential~\cite{Meisinger:2001cq,Fukushima:2003fw,Dumitru:2010mj}; see
Ref.~\cite{Fukushima:2017csk} for
details.
It should be noted that main results presented in this paper are
model independent, qualitatively.
We use the effective model to perform the numerical calculations to
demonstrate our knowledge.

The Lagrangian density of the two-flavor and three-color PNJL model in
the Euclidean space is
\begin{align}
 {\cal L}
 &= {\bar q} (\Slash{D} + m_0)q
  \nonumber\\
 &- G[({\bar q}q)^2+({\bar q}i\gamma_5 {\vec \tau} q)^2]
  + {\cal V}_{\mathrm{g}} (\Phi,{\bar \Phi}),
\end{align}
where $q$ represents the quark field,
the covariant derivative is
$D_\nu=\partial_\nu - i g A_\nu \delta_{\nu 4}$,
${\cal V}_{\mathrm{g}}$ expresses the gluonic contribution
and $\Phi$ (${\bar \Phi}$) means the Polyakov-loop (its conjugate).
The actual form of the effective potential with the mean-field
approximation is
\begin{align}
{\cal V}_{\mathrm{PNJL}}
&= {\cal V}_{\mathrm{f}} + {\cal V}_{\mathrm{g}},
\end{align}
with
\begin{align}
{\cal V}_{\mathrm{f}}
&= - 2 N_\mathrm{f} \int_\Lambda \frac{d^3 p}{(2\pi)^3}
   \Bigl[ N_\mathrm{c} E_{\bf p}
        + T \ln \Bigl( f^- f^+ \Bigr)\Bigr]
   + G \sigma^2,
\end{align}
where $\Lambda$ is the three-dimensional momentum cutoff.
The Fermi-Dirac distribution functions become
\begin{align}
f^-
&= 1
 + 3 (\Phi+{\bar \Phi} e^{-\beta E_{\bf p}^-} ) e^{-\beta E_{\bf p}^-}
 + e^{-3\beta E_{\bf p}^-},~~~~
\nonumber\\
f^+
&= 1
 + 3 ({\bar \Phi}+\Phi e^{-\beta E_{\bf p}^+} ) e^{-\beta E_{\bf p}^+}
 + e^{-3\beta E_{\bf p}^+},
\end{align}
where $E^\mp_{\bf p} = E_{\bf p} \mp \mu$ and $E_{\bf p} = \sqrt{{\bf p}^2 +
M^2}$ with $\mu=(\mu_\mathrm{R},\mu_\mathrm{I})$ and $M = m_0 - 2 G \sigma$.
The condensate $\sigma$ is defined as $\sigma \equiv \langle {\bar q} q
\rangle$.
The parameter set used in the NJL part is taken from
Ref.~\cite{Kashiwa:2007hw}.

In this article, we employ the logarithmic Polyakov-loop effective
potential~\cite{Roessner:2006xn} as the effective model for the gluonic
contribution.
The functional form is
\begin{align}
\frac{{\cal V}_{\mathrm{g}}}{T^4}
 &= - \frac{1}{2} a(T) {\bar \Phi} \Phi
 \nonumber\\
 &  + b(T)
      \ln \Bigl[ 1 - 6{\bar \Phi} \Phi + 4 ({\bar \Phi}^3 + \Phi^3)
               - 3({\bar \Phi}\Phi)^2 \Bigr],
\label{Eq:log_PNJL}
\end{align}
with
\begin{align}
a(T) &= a_0 + a_1 \Bigl( \frac{T_0}{T} \Bigr)
            + a_2 \Bigl( \frac{T_0}{T} \Bigr)^2,
 \nonumber\\
b(T) &= b_3 \Bigl( \frac{T_0}{T} \Bigr)^3,
\end{align}
where parameters, $(a_0,a_1,a_2,b_3)$, are taken from
Ref.~\cite{Roessner:2006xn}.
The remaining parameter $T_0$ is usually fixed as $270$ MeV which is the
critical temperature in the pure gauge limit.
In the case of full QCD, $T_0$ may be determined to reproduce the
pseudo-critical temperature of the deconfinement transition at zero
$\mu$.
However, we still use $T_0=270$ MeV because we are interested in
the qualitative behavior in this article.
The RW endpoint temperature in the present setting is about
$T_\mathrm{RW}=254$ MeV.
In following discussions, we use the PNJL model with above setting.

\section{Canonical ensemble method}
\label{Sec:ce}

\begin{figure}[t]
 \centering
 \includegraphics[width=0.45\textwidth]{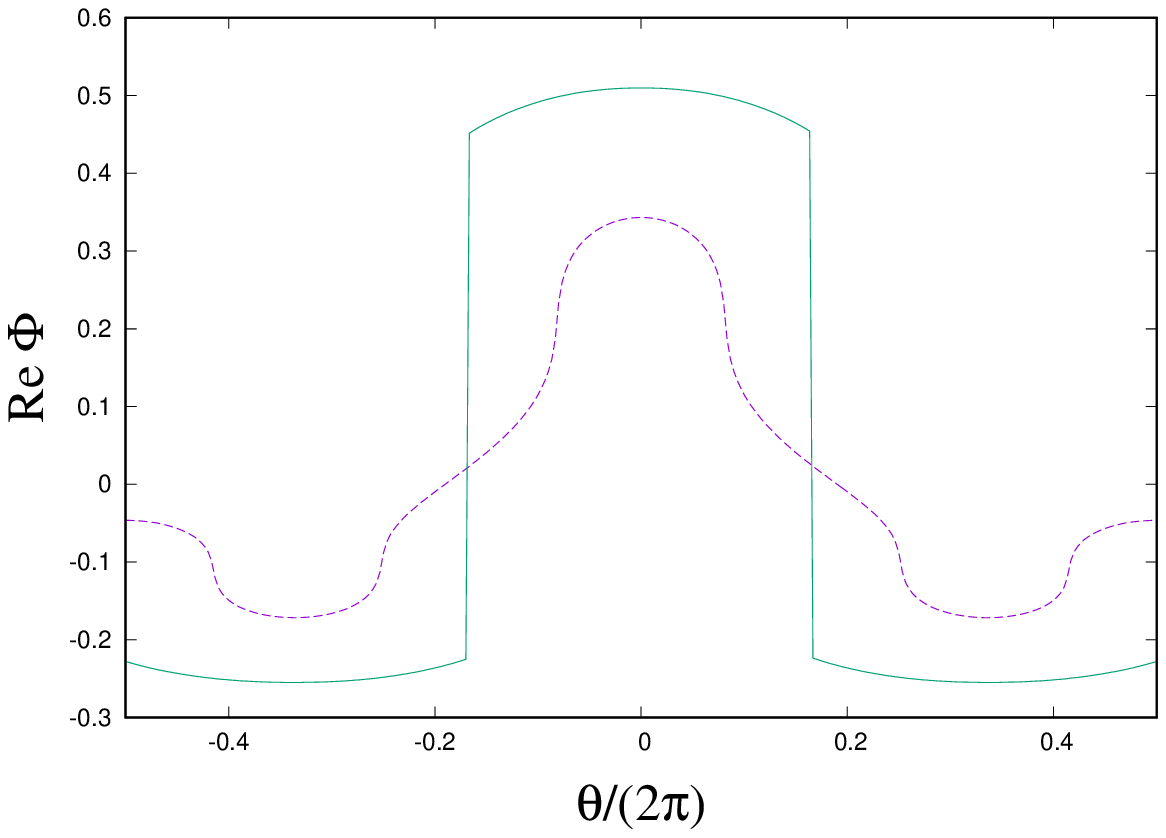}
 \includegraphics[width=0.45\textwidth]{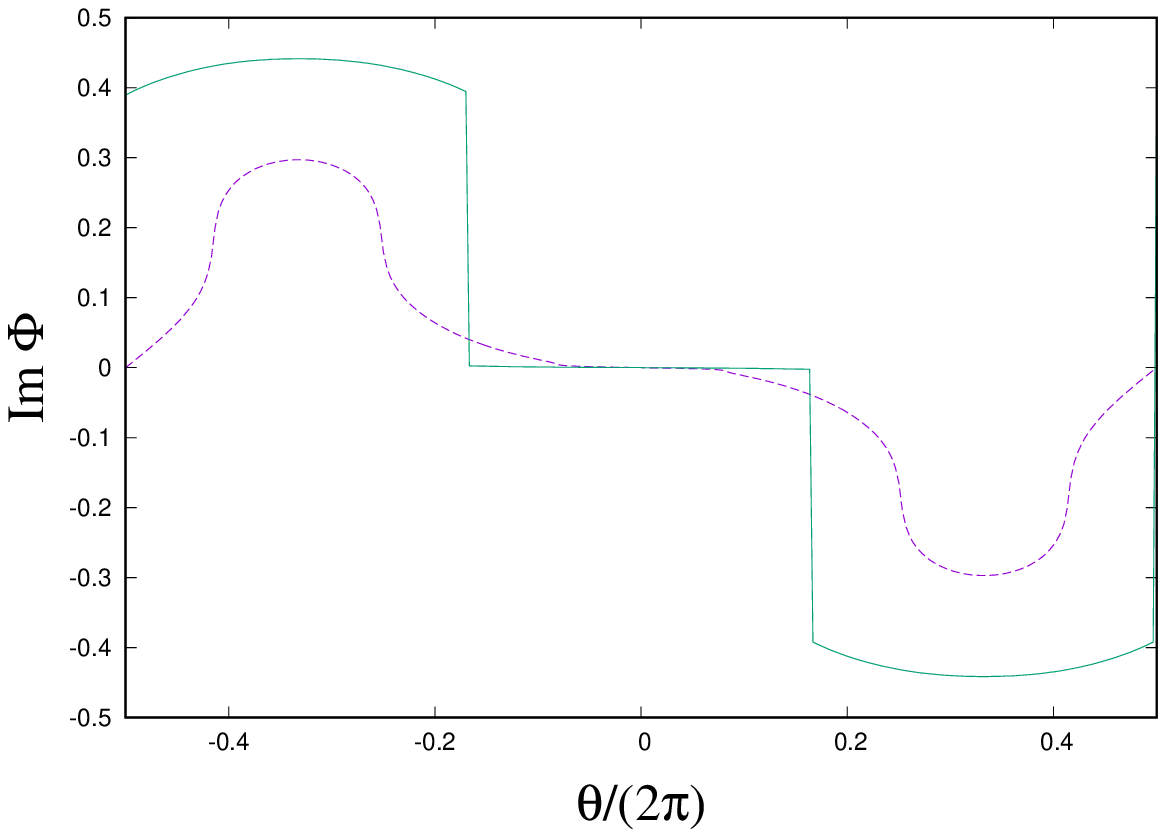}
 \caption{
 The top and bottom panels show the
 $\theta$-dependence of the real and imaginary parts of the Polyakov
 loop, respectively.
 The dotted and solid lines are results at $T=240$ and $260$ MeV,
 respectively.
 }
 \label{Fig:F-q}
\end{figure}

In the standard calculation to investigate the QCD phase diagram,
we start from the grand-canonical partition function;
\begin{align}
 {\cal Z}_\mathrm{GC}(T,\mu)
 &= \int [{\cal D}A][{\cal D}{q}][{\cal D}{\bar q}] e^{-S_\mathrm{QCD}(\mu)},
\end{align}
where $S_\mathrm{QCD}$ is the QCD Euclidean action and $A$ means
the gluon field.
It is well known that we can construct the canonical partition function
with fixed real quark number (Q), ${\cal Z}_\mathrm{C}(T,Q)$, from the
grand-canonical partition function by using
$\theta$~\cite{Roberge:1986mm,Hasenfratz:1991ax} as
\begin{align}
 {\cal Z}_\mathrm{C}(T,Q)
 &= \frac{1}{2 \pi}
    \int_{-\pi}^\pi d \theta~ e^{-i Q \theta}
                             {\cal Z}_\mathrm{GC}(T,\theta).
\label{Eq:CPF}
\end{align}
The domain of integration is usually fixed as $\theta=[-\pi,\pi]$.
Since the QCD grand-canonical partition function has the
RW periodicity, Eq.\,(\ref{Eq:CPF}) can be rewritten as
\begin{align}
  {\cal Z}_\mathrm{C}(T,Q)
 &= {\cal Z}^Q_{C} \Bigl(T,-\pi,-\frac{\pi}{3}          \Bigr)
 + {\cal Z}^Q_{C} \Bigl(T,-\frac{\pi}{3},\frac{\pi}{3} \Bigr)
 \nonumber\\
 &+ {\cal Z}^Q_{C} \Bigl(T, \frac{\pi}{3}, \pi          \Bigr),
\end{align}
where
 \begin{align}
  {\cal Z}^Q_{C}(T,a,b)
  &= \frac{1}{2\pi} \int_{a}^{b}  d \theta~ e^{-i Q \theta}
                                    {\cal Z}_\mathrm{GC}(T,\theta).
 \end{align}
Therefore,  ${\cal Z}_\mathrm{C}(T,Q)$ becomes
 \begin{align}
 {\cal Z}_\mathrm{C}
  &= \Bigl( 1 + e^{ i \frac{2\pi Q}{3}}
              + e^{-i \frac{2\pi Q}{3}} \Bigr)
     {\cal Z}^Q_{C} \Bigl(T,-\frac{\pi}{3},\frac{\pi}{3} \Bigr)
  \nonumber\\
    & = \begin{dcases}
     0 \hspace{2.85cm} Q \neq 0~\mathrm{mod}~3 \\
     3  {\cal Z}^Q_{C} \Bigl(T,-\frac{\pi}{3},\frac{\pi}{3} \Bigr)
     \hspace{0.5cm} Q = 0~\mathrm{mod}~3. \\
	 \end{dcases}
 \end{align}
Then, the expectation value of an arbitrary {\it RW periodic} operator
 (${\cal O}$) is given by
\begin{align}
 \langle {\cal O} \rangle_\mathrm{C} %(T,Q)
 &= \frac{1}{2 \pi}
    \int_{-\pi}^\pi d \theta~
    \Bigl( \frac{{\mathcal Z}_{\mathrm{GC}}}{{\cal Z}_\mathrm{C}}
    \Bigr) e^{-i Q \theta}
                    \langle {\cal O} \rangle_\mathrm{GC}
 \nonumber\\
	 & = \begin{dcases}
	 \hspace{0.285cm}
	 \mathrm{undefined ~or~finite} \hspace{0.8cm} Q \neq 0~\mathrm{mod}~3 \\
%	 \frac{3}{2 \pi}
%         \int_{-\pi/3}^{\pi/3} d \theta~ \frac{{\mathcal
%	     Z}_{\mathrm{G.C.}}}{{\cal Z}_\mathrm{C}} {\cal O}e^{-i Q \theta}
%                               \langle {\cal O} \rangle_\mathrm{G.C.}
         \hspace{0.3cm} \mathrm{can~be~nonzero} ~~~~~~~~~~~ Q = 0~\mathrm{mod}~3.\\
	      \end{dcases}
\label{Eq:cp}
\end{align}
The ``undefined'' means that we encounter $0/0$ in the calculation,
but it can be defined if we remove the factor, $(1+z+z^2)$ where
$z=\exp(2\pi Q i /3)$, by reducing the fractions
to a common denominator because the factor
appears in both the denominator and the numerator.
Below, we follow the latter interpretation; $\langle {\cal O}
\rangle_C$ with $Q \neq 0~\mathrm{mod}~3$ are finite.
This expression cannot be used for the Polyakov loop because
the Polyakov loop is transformed as $e^{2i\pi k/3}\Phi$ under the
$\mathbb{Z}_3$-transformation.
This means that the Polyakov loop does not show the RW periodicity;
it can be clearly seen from Fig.~\ref{Fig:F-q}.
%%%%%%%%%%%%%%%%%%%%%%%%%%%%%%%%%%%%%%%%%%%%%%%%%%%
%%%%%%%%%%%%%%%%%%%%%%%%%%%%%%%%%%%%%%%%%%%%%%%%%%%
Because of the RW transition, $\mathrm{Re}\, \Phi$ and $\mathrm{Im}\, \Phi$
have the gap at $\theta=(2k-1)\pi/3$ at sufficiently high $T$.

In comparison, the expectation value of the Polyakov loop which is the
{\it RW un-periodic} operator
becomes
\begin{align}
 \langle \Phi \rangle_\mathrm{C} %(T,Q)
% \nonumber\\
 &= \frac{1}{2 \pi}
    \int_{-\pi}^\pi d \theta~
    \Bigl( \frac{{\mathcal Z}_{\mathrm{GC}}}{{\cal Z}_\mathrm{C}}
    \Bigr) e^{-i Q \theta}
                    \langle \Phi \rangle_\mathrm{GC}
 \nonumber\\
	 &= \begin{dcases}
         \hspace{0.3cm} \mathrm{undefined}~~~~~ Q \neq 0~\mathrm{mod}~3 \\
\hspace{0.3cm}  0 \hspace{1.9cm} Q= 0~\mathrm{mod}~3.
\\
	    \end{dcases}
\label{Eq:canonical}
\end{align}
Therefore, the Polyakov loop behaves differently comparing with the RW
periodic quantities (\ref{Eq:cp}).
This behavior induces the paradox which is so called the Polyakov-loop
paradox;
the canonical sectors with the quark number $Q \neq 0~\mathrm{mod}~3$
become unphysical and thus we must only sum up contributions with $Q =
0~\mathrm{mod}~3$ to avoid the divergence when we calculate
$\langle \Phi \rangle$ by
using $Z_\mathrm{C}$;
\begin{align}
 \langle \Phi \rangle_\mathrm{GC'} %(T,\mu)
 & = \sum_{Q=-\infty}^{\infty} e^{Q \mu/T}
    \Bigl( \frac{{\mathcal Z}_{\mathrm{C}}}{{\cal Z}_\mathrm{GC}}
    \Bigr) \langle \Phi \rangle_\mathrm{C} (T,Q)
 \nonumber\\
 &\to \sum_{B=-\infty}^{\infty} e^{3B \mu/T}
    \Bigl( \frac{{\mathcal Z}_{\mathrm{C}}}{{\cal Z}_\mathrm{GC}}
    \Bigr) \langle \Phi \rangle_\mathrm{C} (T,3B).
 \label{Eq:La}
\end{align}
Then, we always obtain
\begin{align}
 \langle \Phi \rangle_\mathrm{GC'} %(T,\mu)
 = 0,
\end{align}
from the canonical ensemble method at any $T$.
This is so called the Polyakov loop paradox~\cite{Kratochvila:2006jx}.

 \section{Roles of Modified Polyakov-loop}
\label{Sec:mpl}

To evade the Polyakov-loop paradox, we here consider the modified
Polyakov-loop~\cite{Sakai:2008py} defined as
\begin{align}
\Psi &= e^{i \theta} \Phi.
\end{align}
On the RW transition lines at $\theta=(2k-1)\pi/3$,
the imaginary part of the quark number
density and the modified Polyakov-loop can
have the nonzero value.
This transition can be understood from the spontaneous shift symmetry
($({\mathbb{Z}_2)}_\mathrm{shift}$) breaking.
The $(\mathbb{Z}_2)_\mathrm{shift}$ symmetry at
$\theta=(2\pi-1)/\mathrm{N_c}$ is the invariance under the
transformation associated with the time reversal (${\cal T}$) or
the charge conjugation (${\cal C}$) and
$\mathbb{Z}_\mathrm{N_c}$
transformations via the semidirect
product~\cite{Kashiwa:2012xm,Shimizu:2017asf,kikuchi2018t,Nishimura:2019umw};
\begin{align}
\mathbb{Z}_2 \rtimes \mathbb{Z}_\mathrm{N_c}.
\end{align}
The Polyakov loop under
the $({\mathbb{Z}_2)}_\mathrm{shift}$ transformation just at
$\theta = \lim_{\epsilon \to 0} (\pi/3 - \epsilon)$ is
\begin{align}
 \Phi &= e^{- i\pi/3} |\Phi|
 \xrightarrow[~~\mathbb{Z}_3~]{} e^{ i \pi/3} |\Phi|
 \xrightarrow[~~\mathbb{Z}_2~]{} \Phi~~\mathrm{for~low~} T, \nonumber\\
 \Phi &= |\Phi|
 \xrightarrow[~~\mathbb{Z}_3~]{} e^{ -i 2\pi/3} |\Phi| \xrightarrow[~~\mathbb{Z}_2~]{}  \hspace{-6.5mm} \times  \hspace{4mm} \Phi~~
 \hspace{5.5mm} \mathrm{for~high~} T,
\end{align}
but the imaginary part of the modified
Polyakov-loop is transformed as
\begin{align}
 \mathrm{Im}~\Psi \xrightarrow[({\mathbb{Z}_2)}_\mathrm{shift}]{}
 \hspace{-8mm}\times  \hspace{8mm} \mathrm{Im}~\Psi~~
 \hspace{4mm} \mathrm{for~all~} T,
\end{align}
where $A \xrightarrow[~~{B}~~]{}  \hspace{-8mm}\times \hspace{5mm}C$
means that $A$ is not transformed to $C$ by using $B$ transformation.
Therefore, we can use $\mathrm{Im}\, \Psi$ as the order parameter to detect the
spontaneous $({\mathbb{Z}_2)}_\mathrm{shift}$ symmetry breaking.
In the case of $SU(2)$, we have the stick
symmetry~\cite{Ishiyama:2009bk} and then the
Polyakov loop can be used to detect it exactly at $\theta=(2k-1)\pi/2$.

The modified Polyakov-loop is known as the RW periodic quantity and thus
it can become nonzero value within the canonical ensemble method;
we define it as
\begin{align}
 &\langle \Psi \rangle_\mathrm{C} (Q,T)
 \nonumber\\
 &= \frac{1}{2 \pi}
    \int_{-\pi}^\pi d \theta~
    \Bigl( \frac{{\mathcal Z}_{\mathrm{GC}}}{{\cal Z}_\mathrm{C}}
    \Bigr) e^{-i Q \theta}
                    \langle \Psi \rangle_\mathrm{GC}
 \nonumber\\
	 &= \begin{dcases}
	 \hspace{0.3cm} \mathrm{finite} \hspace{2.48cm} Q \neq 0~\mathrm{mod}~3 \\
         \hspace{0.3cm} \mathrm{can~be~nonzero}~~~~~~~~ Q = 0~\mathrm{mod}~3.\\
	    \end{dcases}
\label{Eq:canonical}
\end{align}
The modified Polyakov-loop can reflect the Polyakov-loop dynamics in QCD
and thus we can evade the Polyakov-loop paradox by using it; see
the next section for more details.
Actually, we can rewrite the Polyakov-loop appearing in the partition
function into the modified Polyakov-loop because $e^{i\theta}$ which can
be interpreted as the fermion boundary condition should be
coupled with the temporal component of the link variable and thus all
dynamical variables in QCD can be RW periodic;
it can be clearly seen in PNJL model in Eq.~(16) of Ref.~\cite{Sakai:2008py}.

\section{Equivalent representation}
\label{Sec:dqc}
At $\mu_\mathrm{R}=0$, we can simply evaluate the expectation
values via the following representation denoted by
$\langle \cdots \rangle_\mathrm{C}'$;
\begin{align}
 \langle {\cal O} \rangle_\mathrm{C}^\prime (T,Q)
 &\equiv  \frac{1}{2 \pi}
    \int_{-\pi}^\pi d \theta~
    e^{-i Q \theta} \langle {\cal O} \rangle_\mathrm{GC},
\label{Eq:pGC}
\end{align}
because we can reach the expression
\begin{align}
 \sum_{Q=-\infty}^\infty \langle {\cal O} \rangle^\prime_\mathrm{C}
 = \sum_{Q=-\infty}^\infty
   \Bigl( \frac{{\cal Z}_\mathrm{C}}{{\cal Z}_\mathrm{GC}(\theta=0)} \Bigr)
   \langle {\cal O} \rangle_\mathrm{C},
\label{Eq:eq}
\end{align}
from following procedure:
This expression is obtained by using the relation
\begin{align}
 \frac{1}{2\pi}
 \sum_{Q=-\infty}^\infty e^{-iQ \theta} &= F(\theta),~~~~
 %2\pi \delta(\theta),~~~~
 -\pi \le \theta \le \pi,
 \label{Eq:D}
\end{align}
where $F(\theta)$ is the $2 \pi$-periodic delta function
\begin{align}
F(\theta) &= \sum_k \delta(\theta + 2\pi k),
\end{align}
here the $2\pi$-periodicity of the Matsubara frequency
is the origin of the periodicity of $F(\theta)$.
Then, the following relation is {\it mathematically} true;
\begin{align}
 &\langle \Phi \rangle^\prime_\mathrm{C} (T,Q)
 \nonumber\\
 %&\equiv  \frac{1}{2 \pi}
 %   \int_{-\pi}^\pi d \theta~
 %   e^{-i Q \theta} \langle \Phi \rangle_\mathrm{GC}
 %\nonumber\\
	 &= \begin{dcases}
%	 \frac{3}{2 \pi}
%         \int_{-\pi/3}^{\pi/3} d \theta~ {\cal P} e^{-i Q \theta}
%                               \langle {\cal O} \rangle_\mathrm{G.C.}
         \hspace{0.3cm} 0 \hspace{2.9cm} Q+1 \neq 0~\mathrm{mod}~3 \\
\hspace{0.3cm}  \mathrm{can~be~nonzero} \hspace{7.5mm} Q+1= 0~\mathrm{mod}~3.
\\
	    \end{dcases}
\label{Eq:canonical_2}
\end{align}
Since the partition function itself is difficult to treat in the effective model with
the mean-field approximation because we should take the thermodynamic limit,
Eq.~(\ref{Eq:pGC}) is convenient for our purpose.
With Eq.~(\ref{Eq:pGC}), we can qualitatively discuss the canonical
ensemble because it shares same properties with $\langle {\cal O}
\rangle_\mathrm{C}$;
for example, the expectation value of the Polyakov-loop in the
grand-canonical ensemble has contributions from the sector with $Q~\neq
0~\mathrm{mod}~3$~\cite{Kratochvila:2006jx}.

To go beyond the Polyakov-loop paradox, we consider
the Fourier transformation of the modified Polyakov-loop;
\begin{align}
 \langle \Psi \rangle^\prime (T,Q)
 &= \frac{1}{2 \pi}
    \int_{-\pi}^\pi d \theta~
    e^{-i Q \theta} \langle \Psi \rangle_\mathrm{GC}
 \nonumber\\
  &= \frac{1}{2 \pi}
    \int_{-\pi}^\pi d \theta~
    e^{-i (Q-1) \theta}
                   \langle \Phi \rangle_\mathrm{GC}.
\end{align}
The most important point in this article is that
we can have $\langle \Psi \rangle = \langle \Phi \rangle$ at
$\mu_\mathrm{R}=0$ by shifting $(Q-1)$ to $Q'$ as
\begin{align}
 \langle \Phi \rangle_\mathrm{GC}
 = \sum_{Q = -\infty}^{\infty} \langle \Phi \rangle'(T,Q)
 &= \sum_{Q' = -\infty}^{\infty} \langle \Psi \rangle'(T,Q)
 = \langle \Psi \rangle_\mathrm{GC},
\end{align}
when $Q$ and $Q'$ run from $-\infty$ to $\infty$.
This relation means that
{\it we can evade the Polyakov-loop paradox} by using
the modified Polyakov-loop when  we calculate the Polyakov-loop with the
canonical ensemble because $\Psi$ is the  RW periodic quantity.
There may be the difference between
$\langle \Phi \rangle$
and $\langle \Psi \rangle$ in the finite size system where maximum $|Q|$
and $|Q'|$ exist.
It should be noted that we cannot exactly discuss $\langle \Phi
\rangle_\mathrm{GC}$ at $\mu_\mathrm{R}\neq 0$ via
$\langle \Psi \rangle_\mathrm{C}'$ because we use Eq.\,(\ref{Eq:D}) with
the fugacity expansion, but
it is possible to compute $\langle \Phi \rangle_\mathrm{GC}$
via $\langle \Psi \rangle_\mathrm{C}$, in principle.

\begin{figure}[t]
 \centering
 \includegraphics[width=0.45\textwidth]{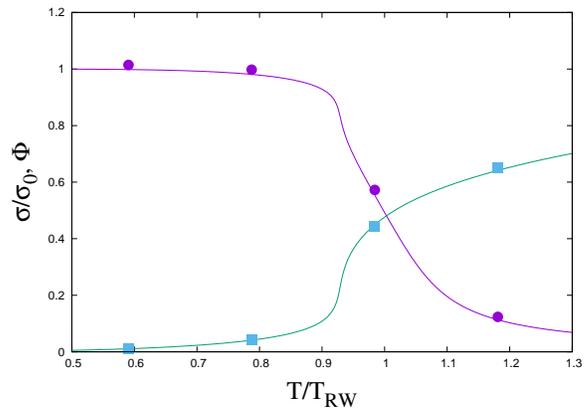}
 \caption{
 The $T$-dependence of the chiral condensate and the Polyakov loop,
 receptively.
 The solid lines are results calculated from the grand-canonical
 partition function and the symbols are results calculated from
 Eq.~(\ref{Eq:pGC}) with the fugacity expansion.
 The chiral condensate is normalized by that at $T=\mu=0$.
 }
 \label{Fig:sigma_Polyakovloop}
\end{figure}

In addition, above fact indicates that we can restrict the domain of
integral to $-\pi/3 \sim \pi/3$ from $-\pi \sim \pi$ in the Fourier
transformation because its domain of integral can be restricted in
one period;
\begin{align}
  \Sigma^{(n)}
 :&= \frac{1}{2 \pi}
    \int_{-\pi/3}^{\pi/3} d \theta~ e^{-i n \theta}
 \langle \sigma \rangle_\mathrm{GC},
 \\
 \Psi^{(n)}
 :&= \frac{1}{2 \pi}
    \int_{-\pi/3}^{\pi/3} d \theta~ e^{-i n \theta}
 \langle \Psi \rangle_\mathrm{GC},
 \nonumber\\
 &= \frac{1}{2 \pi}
    \int_{-\pi/3}^{\pi/3} d \theta~ e^{-i (n-1) \theta}
 \langle \Phi \rangle_\mathrm{GC}
 :=   \Phi^{(n-1)},
\label{Eq:dq}
\end{align}
This expression indicates that the non-trivial $\mathbb{Z}_3$-images
do not have any clear physical meaning.
To check this restriction can work or not, we numerical evaluate the chiral
condensate and the Polyakov loop in Fig.~\ref{Fig:sigma_Polyakovloop}.
%%%%%%%%%%%%%%%%%%%%%%%%%%%%%%%%%%%%%%%%%%%%%%%%%%%
%%%%%%%%%%%%%%%%%%%%%%%%%%%%%%%%%%%%%%%%%%%%%%%%%%%
The solid lines are results directly calculated from the grand-canonical
partition function and the symbols are results calculated from
Eq.~(\ref{Eq:dq}).
Since the convergence against $Q$ is rather
slow, the symbol and the solid line are slightly different, but it will be
matched with each other when we take into account sufficient number of
$Q$.
Interestingly, we can exactly reproduce the chiral condensate and also
the Polyakov loop by using the restriction of the integral range.
This indicates that the non-trivial $\mathbb{Z}_3$ images are
unphysical and then we can remove it from the Fourier transformation.
This means that the real chemical potential region which is belonging to the
trivial $\mathbb{Z}_3$ sector is constructed from the trivial
$\mathbb{Z}_3$ sector at finite imaginary chemical potential.

\begin{figure}[t]
 \centering
 \includegraphics[width=0.45\textwidth]{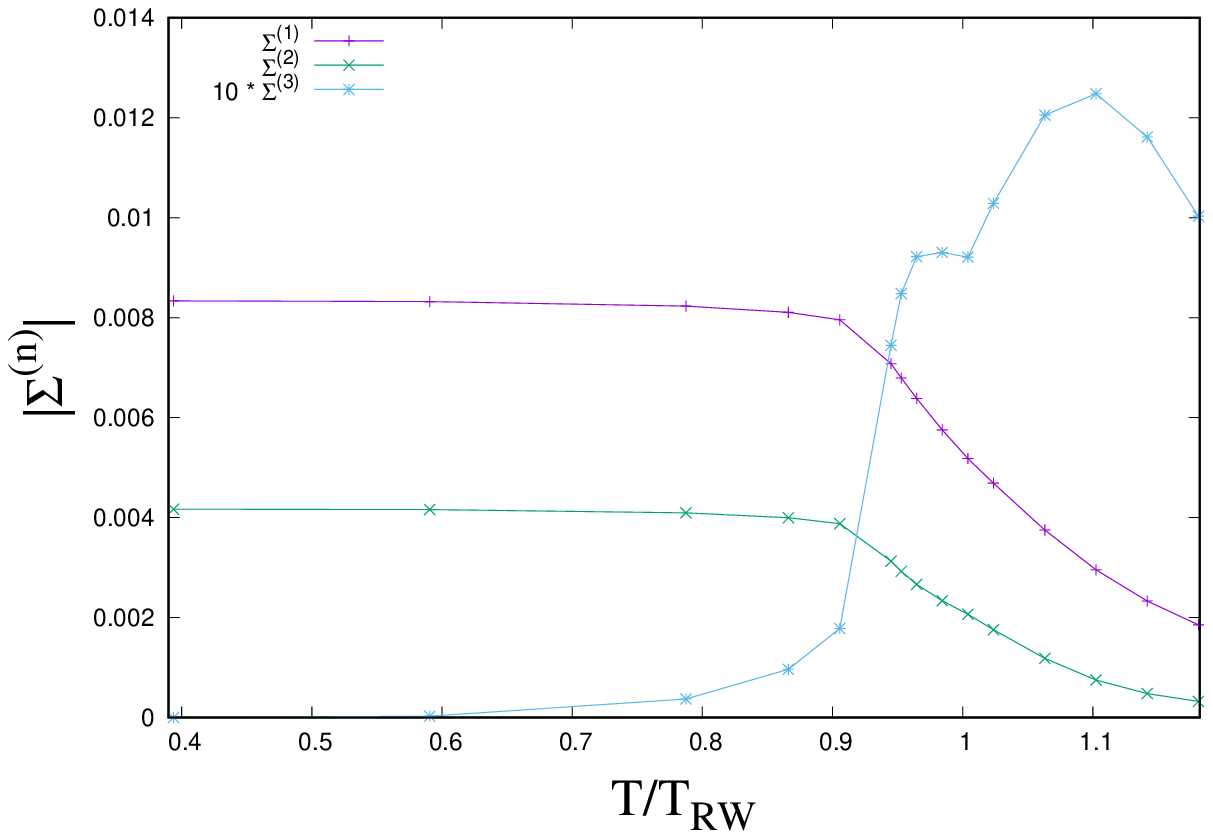}
 \includegraphics[width=0.45\textwidth]{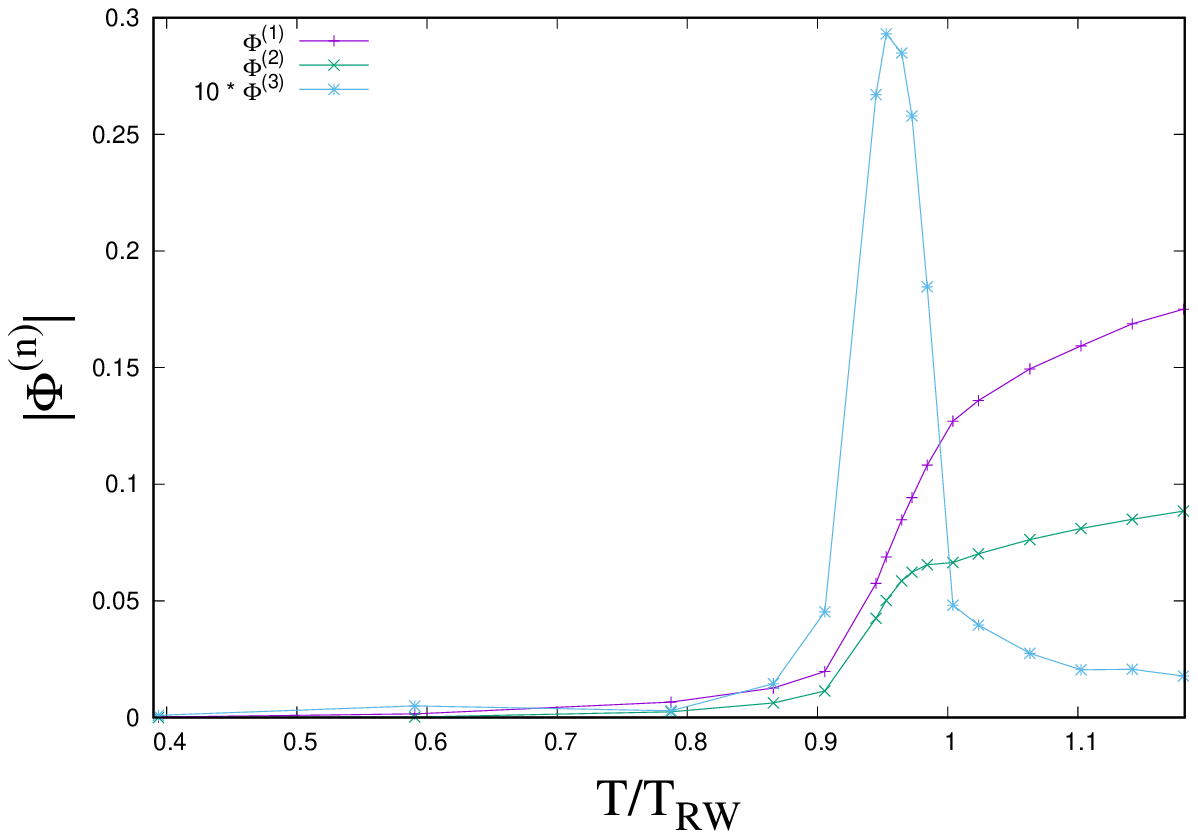}
 \caption{
 The $T$-dependence of dual quark with $n=1$, $2$ and $3$.
 }
 \label{Fig:dq}
\end{figure}

The chiral condensate under the restriction of the integral (\ref{Eq:dq})
seems to be similar to the dual quark
condensate~\cite{Bilgici:2008qy} defined
as
\begin{align}
 \Sigma^{(n)}
 &= -\int_{-\pi}^\pi \frac{d \varphi}{2\pi}~
 e^{-in\varphi} \sigma(\varphi),
\end{align}
where $\varphi$ is the phase of the quark boundary condition and thus
$\varphi=\theta+\pi$ because the imaginary chemical potential can be
converted to the quark boundary condition~\cite{Kashiwa:2009ki}.
The dual quark condensate with $n \neq 0~\mathrm{mod}~3$ has been
considered as the order parameter or the indicator of the
confinement-deconfinement transition.
This quantity has been investigated by using the lattice QCD
simulation~\cite{Bilgici:2008qy,Bilgici:2009tx,*Bilgici:2009phd,Bruckmann:2011zx},
the Dyson-Schwinger equations~\cite{Fischer:2009wc,*Fischer:2009gk}, the PNJL
model~\cite{Kashiwa:2009ki,Gatto:2010qs,Zhang:2015baa,*Zhang:2017bem} and so on.
From the same reason that the canonical partition function
with $Q \neq 0~\mathrm{mod}~3$ becomes zero,
we must break the RW periodicity to compute the dual chiral condensate
with $n \neq 0~\mathrm{mod}~3$ when the system contains the
dynamical quarks.
The ordinary method to break the RW periodicity is that the gauge
configuration is fixed at the anti-periodic quark boundary
condition and it is
used to another value of
$\varphi$~\cite{Bilgici:2009tx,Bilgici:2009phd}, but there is no
justification of the method so far;
we do not need this artificial procedure in the quenched limit.
However, when we accept the restriction of the integral as
Eqs.\,(\ref{Eq:dq}),
the RW periodicity is not the matter and then we can well determine the
dual quark condensate; it is nothing but each canonical sector of the
chiral condensate calculated from the grand-canonical ensemble with
$\mu_\mathrm{I}$ via the
Fourier transformation.
From this viewpoint, we can expect that the Fourier components of the
chiral condensate and also the Polyakov loop have some hints to obtain
the deeper understanding of the QCD properties, particularly the
confinement-deconfinement transition.
The actual behavior of the absolute value of the dual quark-condensate
and Polyakov-loop components with $n=1$,
$2$ and $3$ are shown in Fig.~\ref{Fig:dq}.
%%%%%%%%%%%%%%%%%%%%%%%%%%%%%%%%%%%%%%%%%%%%%%%%%%%
%%%%%%%%%%%%%%%%%%%%%%%%%%%%%%%%%%%%%%%%%%%%%%%%%%%
Interestingly, $\Sigma^{(3)}$ behaves as similar with the Polyakov
loop unlike the ordinary dual quark condensate.
Therefore, we can use $\Sigma^{(3)}$ as the order parameter or the
indicator of the confinement-deconfinement transition under the
restriction of the integral range.
Also, $n=3$ component shows non-monotonic behavior under the
restriction;
when we approach the RW endpoint temperature, $\Phi^{(3)}$ quickly
approaches to small value.

\section{Summary}
\label{Sec:Summary}

In this article, we have investigated the meaning of the
$\mathbb{Z}_{N_\mathrm{c}}$ images appearing in QCD at finite
temperature ($T$) and imaginary
chemical potential ($\mu_\mathrm{I}$).
Particularly, to answer how the $\mathbb{Z}_{N_\mathrm{c}}$ images affect the
thermal system, we have considered the canonical ensemble by using the
grand-canonical ensemble via the Fourier transformation.
In actual discussions, we employ the different representation of the
expectation value via the Fourier transformation % and the fugacity expansion
which is mathematically equivalent with the canonical
ensemble method at zero real chemical potential; see Eq.~(\ref{Eq:eq}).

When we integrate out the grand-canonical partition function,
${\cal Z}_\mathrm{GC} (T,\mu_\mathrm{I})$, from
$\theta=\mu_\mathrm{I}/T=-\pi$ to $\pi$ to construct the canonical
partition function with fixed quark number ($Q$),
only the $Q=0~\mathrm{mod}~3$ sector contributes to the
thermodynamic system.
However, the Polyakov-loop paradox appears in this case;
the expectation value of the Polyakov loop by using the canonical
ensemble becomes infinity or $0$ after removing the unphysical sectors
and it is inconsistent with the value
calculated from the grand-canonical ensemble.
This paradox is coursed from the fact that the Polyakov loop does
not have the Roberge-Weiss periodicity which is the
special property of QCD grand-canonical partition function at finite
$\theta:=\mu_\mathrm{I}/T$.

To discuss the Polyakov-loop paradox, we revisit the modified
Polyakov-loop.
Then, the expectation value of the Polyakov loop ($\Phi$)
can be
constructed from the modified Polyakov-loop ($\Psi=e^{i \theta} \Phi$)
which is the RW periodic quantity.
In the case of the RW periodic quantities, only the $Q=0~\mathrm{mod}~3$
sector contributes the thermodynamic system and then we can prove the relation,
$\langle \Psi \rangle_\mathrm{GC} = \langle \Phi \rangle_\mathrm{GC}$,
at least with zero real chemical potential.
Thus, there is no Polyakov-loop
paradox.

From the RW periodicity issue of the modified Polyakov-loop,
we can restrict the integral range from $-\pi \sim \pi$ to $-\pi/3 \sim
\pi/3$ because of the property of the Fourier transformation.
Actually, we can correctly reproduce the chiral condensate and also
the Polyakov loop by using this restriction.
This result strongly indicates that the non-trivial
$\mathbb{Z}_3$ images are unphysical and we must remove it
from the Fourier transformation when we construct the canonical ensemble.
Following the restriction,
we can well determine the dual quark condensate;
in the ordinary determination, we must break the RW periodicity by hand,
but now we do not need such unclear procedure.
We have demonstrated the $T$-dependence of each canonical
sector of the chiral condensate and the Polyakov loop by using the PNJL model
under the restriction of the integral range.

Recently, it is proposed that the structure of QCD at finite
$\mu_\mathrm{I}$ may be related with the confinement-deconfinement
transition~\cite{Kashiwa:2015tna,*Kashiwa:2016vrl,*Kashiwa:2017yvy} based
on the analogy of the topological order at
$T=0$~\cite{Wen:1989iv,Sato:2007xc}
and thus it is interesting to know detailed properties of QCD at finite
$\mu_\mathrm{I}$.
We hope that this analysis shed a light to mysterious property of the
confinement-deconfinement transition.

\begin{acknowledgments}
 K.K. and H.K. are supported by the Grants-in-Aid for Scientific Research
 from JSPS (No. 18K03618) and (No. 17K05446), respectively.
\end{acknowledgments}

\bibliography{ref.bib}

\begin{thebibliography}{53}%
\makeatletter
\providecommand \@ifxundefined [1]{%
 \@ifx{#1\undefined}
}%
\providecommand \@ifnum [1]{%
 \ifnum #1\expandafter \@firstoftwo
 \else \expandafter \@secondoftwo
 \fi
}%
\providecommand \@ifx [1]{%
 \ifx #1\expandafter \@firstoftwo
 \else \expandafter \@secondoftwo
 \fi
}%
\providecommand \natexlab [1]{#1}%
\providecommand \enquote  [1]{``#1''}%
\providecommand \bibnamefont  [1]{#1}%
\providecommand \bibfnamefont [1]{#1}%
\providecommand \citenamefont [1]{#1}%
\providecommand \href@noop [0]{\@secondoftwo}%
\providecommand \href [0]{\begingroup \@sanitize@url \@href}%
\providecommand \@href[1]{\@@startlink{#1}\@@href}%
\providecommand \@@href[1]{\endgroup#1\@@endlink}%
\providecommand \@sanitize@url [0]{\catcode `\\12\catcode `\$12\catcode
  `\&12\catcode `\#12\catcode `\^12\catcode `\_12\catcode `\%12\relax}%
\providecommand \@@startlink[1]{}%
\providecommand \@@endlink[0]{}%
\providecommand \url  [0]{\begingroup\@sanitize@url \@url }%
\providecommand \@url [1]{\endgroup\@href {#1}{\urlprefix }}%
\providecommand \urlprefix  [0]{URL }%
\providecommand \Eprint [0]{\href }%
\providecommand \doibase [0]{http://dx.doi.org/}%
\providecommand \selectlanguage [0]{\@gobble}%
\providecommand \bibinfo  [0]{\@secondoftwo}%
\providecommand \bibfield  [0]{\@secondoftwo}%
\providecommand \translation [1]{[#1]}%
\providecommand \BibitemOpen [0]{}%
\providecommand \bibitemStop [0]{}%
\providecommand \bibitemNoStop [0]{.\EOS\space}%
\providecommand \EOS [0]{\spacefactor3000\relax}%
\providecommand \BibitemShut  [1]{\csname bibitem#1\endcsname}%
\let\auto@bib@innerbib\@empty
%</preamble>
\bibitem [{\citenamefont {de~Forcrand}(2009)}]{deForcrand:2010ys}%
  \BibitemOpen
  \bibfield  {author} {\bibinfo {author} {\bibfnamefont {P.}~\bibnamefont
  {de~Forcrand}},\ }\href@noop {} {\bibfield  {journal} {\bibinfo  {journal}
  {PoS}\ }\textbf {\bibinfo {volume} {LAT2009}},\ \bibinfo {pages} {010}
  (\bibinfo {year} {2009})},\ \Eprint {http://arxiv.org/abs/1005.0539}
  {arXiv:1005.0539 [hep-lat]} \BibitemShut {NoStop}%
%%CITATION = ARXIV:1005.0539;%%
\bibitem [{\citenamefont {Allton}\ \emph {et~al.}(2005)\citenamefont {Allton},
  \citenamefont {Doring}, \citenamefont {Ejiri}, \citenamefont {Hands},
  \citenamefont {Kaczmarek} \emph {et~al.}}]{Allton:2005gk}%
  \BibitemOpen
  \bibfield  {author} {\bibinfo {author} {\bibfnamefont {C.}~\bibnamefont
  {Allton}}, \bibinfo {author} {\bibfnamefont {M.}~\bibnamefont {Doring}},
  \bibinfo {author} {\bibfnamefont {S.}~\bibnamefont {Ejiri}}, \bibinfo
  {author} {\bibfnamefont {S.}~\bibnamefont {Hands}}, \bibinfo {author}
  {\bibfnamefont {O.}~\bibnamefont {Kaczmarek}},  \emph {et~al.},\ }\href
  {\doibase 10.1103/PhysRevD.71.054508} {\bibfield  {journal} {\bibinfo
  {journal} {Phys.Rev.}\ }\textbf {\bibinfo {volume} {D71}},\ \bibinfo {pages}
  {054508} (\bibinfo {year} {2005})},\ \Eprint
  {http://arxiv.org/abs/hep-lat/0501030} {arXiv:hep-lat/0501030 [hep-lat]}
  \BibitemShut {NoStop}%
%%CITATION = HEP-LAT/0501030;%%
\bibitem [{\citenamefont {Gavai}\ and\ \citenamefont
  {Gupta}(2008)}]{Gavai:2008zr}%
  \BibitemOpen
  \bibfield  {author} {\bibinfo {author} {\bibfnamefont {R.}~\bibnamefont
  {Gavai}}\ and\ \bibinfo {author} {\bibfnamefont {S.}~\bibnamefont {Gupta}},\
  }\href {\doibase 10.1103/PhysRevD.78.114503} {\bibfield  {journal} {\bibinfo
  {journal} {Phys.Rev.}\ }\textbf {\bibinfo {volume} {D78}},\ \bibinfo {pages}
  {114503} (\bibinfo {year} {2008})},\ \Eprint {http://arxiv.org/abs/0806.2233}
  {arXiv:0806.2233 [hep-lat]} \BibitemShut {NoStop}%
%%CITATION = ARXIV:0806.2233;%%
\bibitem [{\citenamefont {Fodor}\ and\ \citenamefont
  {Katz}(2002{\natexlab{a}})}]{Fodor:2001au}%
  \BibitemOpen
  \bibfield  {author} {\bibinfo {author} {\bibfnamefont {Z.}~\bibnamefont
  {Fodor}}\ and\ \bibinfo {author} {\bibfnamefont {S.}~\bibnamefont {Katz}},\
  }\href {\doibase 10.1016/S0370-2693(02)01583-6} {\bibfield  {journal}
  {\bibinfo  {journal} {Phys.Lett.}\ }\textbf {\bibinfo {volume} {B534}},\
  \bibinfo {pages} {87} (\bibinfo {year} {2002}{\natexlab{a}})},\ \Eprint
  {http://arxiv.org/abs/hep-lat/0104001} {arXiv:hep-lat/0104001 [hep-lat]}
  \BibitemShut {NoStop}%
%%CITATION = HEP-LAT/0104001;%%
\bibitem [{\citenamefont {Fodor}\ and\ \citenamefont
  {Katz}(2002{\natexlab{b}})}]{Fodor:2001pe}%
  \BibitemOpen
  \bibfield  {author} {\bibinfo {author} {\bibfnamefont {Z.}~\bibnamefont
  {Fodor}}\ and\ \bibinfo {author} {\bibfnamefont {S.}~\bibnamefont {Katz}},\
  }\href {\doibase 10.1088/1126-6708/2002/03/014} {\bibfield  {journal}
  {\bibinfo  {journal} {JHEP}\ }\textbf {\bibinfo {volume} {0203}},\ \bibinfo
  {pages} {014} (\bibinfo {year} {2002}{\natexlab{b}})},\ \Eprint
  {http://arxiv.org/abs/hep-lat/0106002} {arXiv:hep-lat/0106002 [hep-lat]}
  \BibitemShut {NoStop}%
%%CITATION = HEP-LAT/0106002;%%
\bibitem [{\citenamefont {Fodor}\ and\ \citenamefont
  {Katz}(2004)}]{Fodor:2004nz}%
  \BibitemOpen
  \bibfield  {author} {\bibinfo {author} {\bibfnamefont {Z.}~\bibnamefont
  {Fodor}}\ and\ \bibinfo {author} {\bibfnamefont {S.}~\bibnamefont {Katz}},\
  }\href {\doibase 10.1088/1126-6708/2004/04/050} {\bibfield  {journal}
  {\bibinfo  {journal} {JHEP}\ }\textbf {\bibinfo {volume} {0404}},\ \bibinfo
  {pages} {050} (\bibinfo {year} {2004})},\ \Eprint
  {http://arxiv.org/abs/hep-lat/0402006} {arXiv:hep-lat/0402006 [hep-lat]}
  \BibitemShut {NoStop}%
%%CITATION = HEP-LAT/0402006;%%
\bibitem [{\citenamefont {Fodor}\ \emph {et~al.}(2003)\citenamefont {Fodor},
  \citenamefont {Katz},\ and\ \citenamefont {Szabo}}]{Fodor:2002km}%
  \BibitemOpen
  \bibfield  {author} {\bibinfo {author} {\bibfnamefont {Z.}~\bibnamefont
  {Fodor}}, \bibinfo {author} {\bibfnamefont {S.}~\bibnamefont {Katz}}, \ and\
  \bibinfo {author} {\bibfnamefont {K.}~\bibnamefont {Szabo}},\ }\href
  {\doibase 10.1016/j.physletb.2003.06.011} {\bibfield  {journal} {\bibinfo
  {journal} {Phys.Lett.}\ }\textbf {\bibinfo {volume} {B568}},\ \bibinfo
  {pages} {73} (\bibinfo {year} {2003})},\ \Eprint
  {http://arxiv.org/abs/hep-lat/0208078} {arXiv:hep-lat/0208078 [hep-lat]}
  \BibitemShut {NoStop}%
%%CITATION = HEP-LAT/0208078;%%
\bibitem [{\citenamefont {de~Forcrand}\ and\ \citenamefont
  {Philipsen}(2002)}]{deForcrand:2002hgr}%
  \BibitemOpen
  \bibfield  {author} {\bibinfo {author} {\bibfnamefont {P.}~\bibnamefont
  {de~Forcrand}}\ and\ \bibinfo {author} {\bibfnamefont {O.}~\bibnamefont
  {Philipsen}},\ }\href {\doibase 10.1016/S0550-3213(02)00626-0} {\bibfield
  {journal} {\bibinfo  {journal} {Nucl. Phys.}\ }\textbf {\bibinfo {volume}
  {B642}},\ \bibinfo {pages} {290} (\bibinfo {year} {2002})},\ \Eprint
  {http://arxiv.org/abs/hep-lat/0205016} {arXiv:hep-lat/0205016 [hep-lat]}
  \BibitemShut {NoStop}%
%%CITATION = HEP-LAT/0205016;%%
\bibitem [{\citenamefont {de~Forcrand}\ and\ \citenamefont
  {Philipsen}(2003)}]{deForcrand:2003vyj}%
  \BibitemOpen
  \bibfield  {author} {\bibinfo {author} {\bibfnamefont {P.}~\bibnamefont
  {de~Forcrand}}\ and\ \bibinfo {author} {\bibfnamefont {O.}~\bibnamefont
  {Philipsen}},\ }\href {\doibase 10.1016/j.nuclphysb.2003.09.005} {\bibfield
  {journal} {\bibinfo  {journal} {Nucl. Phys.}\ }\textbf {\bibinfo {volume}
  {B673}},\ \bibinfo {pages} {170} (\bibinfo {year} {2003})},\ \Eprint
  {http://arxiv.org/abs/hep-lat/0307020} {arXiv:hep-lat/0307020 [hep-lat]}
  \BibitemShut {NoStop}%
%%CITATION = HEP-LAT/0307020;%%
\bibitem [{\citenamefont {D'Elia}\ and\ \citenamefont
  {Lombardo}(2003)}]{DElia:2002tig}%
  \BibitemOpen
  \bibfield  {author} {\bibinfo {author} {\bibfnamefont {M.}~\bibnamefont
  {D'Elia}}\ and\ \bibinfo {author} {\bibfnamefont {M.-P.}\ \bibnamefont
  {Lombardo}},\ }\href {\doibase 10.1103/PhysRevD.67.014505} {\bibfield
  {journal} {\bibinfo  {journal} {Phys. Rev.}\ }\textbf {\bibinfo {volume}
  {D67}},\ \bibinfo {pages} {014505} (\bibinfo {year} {2003})},\ \Eprint
  {http://arxiv.org/abs/hep-lat/0209146} {arXiv:hep-lat/0209146 [hep-lat]}
  \BibitemShut {NoStop}%
%%CITATION = HEP-LAT/0209146;%%
\bibitem [{\citenamefont {D'Elia}\ and\ \citenamefont
  {Lombardo}(2004)}]{DElia:2004ani}%
  \BibitemOpen
  \bibfield  {author} {\bibinfo {author} {\bibfnamefont {M.}~\bibnamefont
  {D'Elia}}\ and\ \bibinfo {author} {\bibfnamefont {M.~P.}\ \bibnamefont
  {Lombardo}},\ }\href {\doibase 10.1103/PhysRevD.70.074509} {\bibfield
  {journal} {\bibinfo  {journal} {Phys. Rev.}\ }\textbf {\bibinfo {volume}
  {D70}},\ \bibinfo {pages} {074509} (\bibinfo {year} {2004})},\ \Eprint
  {http://arxiv.org/abs/hep-lat/0406012} {arXiv:hep-lat/0406012 [hep-lat]}
  \BibitemShut {NoStop}%
%%CITATION = HEP-LAT/0406012;%%
\bibitem [{\citenamefont {Chen}\ and\ \citenamefont {Luo}(2005)}]{Chen:2004tb}%
  \BibitemOpen
  \bibfield  {author} {\bibinfo {author} {\bibfnamefont {H.-S.}\ \bibnamefont
  {Chen}}\ and\ \bibinfo {author} {\bibfnamefont {X.-Q.}\ \bibnamefont {Luo}},\
  }\href {\doibase 10.1103/PhysRevD.72.034504} {\bibfield  {journal} {\bibinfo
  {journal} {Phys.Rev.}\ }\textbf {\bibinfo {volume} {D72}},\ \bibinfo {pages}
  {034504} (\bibinfo {year} {2005})},\ \Eprint
  {http://arxiv.org/abs/hep-lat/0411023} {arXiv:hep-lat/0411023 [hep-lat]}
  \BibitemShut {NoStop}%
%%CITATION = HEP-LAT/0411023;%%
\bibitem [{\citenamefont {Hasenfratz}\ and\ \citenamefont
  {Toussaint}(1992)}]{Hasenfratz:1991ax}%
  \BibitemOpen
  \bibfield  {author} {\bibinfo {author} {\bibfnamefont {A.}~\bibnamefont
  {Hasenfratz}}\ and\ \bibinfo {author} {\bibfnamefont {D.}~\bibnamefont
  {Toussaint}},\ }\href {\doibase 10.1016/0550-3213(92)90247-9} {\bibfield
  {journal} {\bibinfo  {journal} {Nucl. Phys.}\ }\textbf {\bibinfo {volume}
  {B371}},\ \bibinfo {pages} {539} (\bibinfo {year} {1992})}\BibitemShut
  {NoStop}%
%%CITATION = NUPHA,B371,539;%%
\bibitem [{\citenamefont {Alexandru}\ \emph {et~al.}(2005)\citenamefont
  {Alexandru}, \citenamefont {Faber}, \citenamefont {Horvath},\ and\
  \citenamefont {Liu}}]{Alexandru:2005ix}%
  \BibitemOpen
  \bibfield  {author} {\bibinfo {author} {\bibfnamefont {A.}~\bibnamefont
  {Alexandru}}, \bibinfo {author} {\bibfnamefont {M.}~\bibnamefont {Faber}},
  \bibinfo {author} {\bibfnamefont {I.}~\bibnamefont {Horvath}}, \ and\
  \bibinfo {author} {\bibfnamefont {K.-F.}\ \bibnamefont {Liu}},\ }\href
  {\doibase 10.1103/PhysRevD.72.114513} {\bibfield  {journal} {\bibinfo
  {journal} {Phys.Rev.}\ }\textbf {\bibinfo {volume} {D72}},\ \bibinfo {pages}
  {114513} (\bibinfo {year} {2005})},\ \Eprint
  {http://arxiv.org/abs/hep-lat/0507020} {arXiv:hep-lat/0507020 [hep-lat]}
  \BibitemShut {NoStop}%
%%CITATION = HEP-LAT/0507020;%%
\bibitem [{\citenamefont {Kratochvila}\ and\ \citenamefont
  {de~Forcrand}(2006)}]{Kratochvila:2006jx}%
  \BibitemOpen
  \bibfield  {author} {\bibinfo {author} {\bibfnamefont {S.}~\bibnamefont
  {Kratochvila}}\ and\ \bibinfo {author} {\bibfnamefont {P.}~\bibnamefont
  {de~Forcrand}},\ }\href {\doibase 10.1103/PhysRevD.73.114512} {\bibfield
  {journal} {\bibinfo  {journal} {Phys.Rev.}\ }\textbf {\bibinfo {volume}
  {D73}},\ \bibinfo {pages} {114512} (\bibinfo {year} {2006})},\ \Eprint
  {http://arxiv.org/abs/hep-lat/0602005} {arXiv:hep-lat/0602005 [hep-lat]}
  \BibitemShut {NoStop}%
%%CITATION = HEP-LAT/0602005;%%
\bibitem [{\citenamefont {de~Forcrand}\ and\ \citenamefont
  {Kratochvila}(2006)}]{deForcrand:2006ec}%
  \BibitemOpen
  \bibfield  {author} {\bibinfo {author} {\bibfnamefont {P.}~\bibnamefont
  {de~Forcrand}}\ and\ \bibinfo {author} {\bibfnamefont {S.}~\bibnamefont
  {Kratochvila}},\ }\href {\doibase 10.1016/j.nuclphysbps.2006.01.007}
  {\bibfield  {journal} {\bibinfo  {journal} {Nucl.Phys.Proc.Suppl.}\ }\textbf
  {\bibinfo {volume} {153}},\ \bibinfo {pages} {62} (\bibinfo {year} {2006})},\
  \Eprint {http://arxiv.org/abs/hep-lat/0602024} {arXiv:hep-lat/0602024
  [hep-lat]} \BibitemShut {NoStop}%
%%CITATION = HEP-LAT/0602024;%%
\bibitem [{\citenamefont {Li}\ \emph {et~al.}(2010)\citenamefont {Li},
  \citenamefont {Alexandru}, \citenamefont {Liu},\ and\ \citenamefont
  {Meng}}]{Li:2010qf}%
  \BibitemOpen
  \bibfield  {author} {\bibinfo {author} {\bibfnamefont {A.}~\bibnamefont
  {Li}}, \bibinfo {author} {\bibfnamefont {A.}~\bibnamefont {Alexandru}},
  \bibinfo {author} {\bibfnamefont {K.-F.}\ \bibnamefont {Liu}}, \ and\
  \bibinfo {author} {\bibfnamefont {X.}~\bibnamefont {Meng}},\ }\href {\doibase
  10.1103/PhysRevD.82.054502} {\bibfield  {journal} {\bibinfo  {journal}
  {Phys.Rev.}\ }\textbf {\bibinfo {volume} {D82}},\ \bibinfo {pages} {054502}
  (\bibinfo {year} {2010})},\ \Eprint {http://arxiv.org/abs/1005.4158}
  {arXiv:1005.4158 [hep-lat]} \BibitemShut {NoStop}%
%%CITATION = ARXIV:1005.4158;%%
\bibitem [{\citenamefont {Kashiwa}\ and\ \citenamefont
  {Ohnishi}(2015)}]{Kashiwa:2015tna}%
  \BibitemOpen
  \bibfield  {author} {\bibinfo {author} {\bibfnamefont {K.}~\bibnamefont
  {Kashiwa}}\ and\ \bibinfo {author} {\bibfnamefont {A.}~\bibnamefont
  {Ohnishi}},\ }\href {\doibase 10.1016/j.physletb.2015.09.036} {\bibfield
  {journal} {\bibinfo  {journal} {Phys. Lett.}\ }\textbf {\bibinfo {volume}
  {B750}},\ \bibinfo {pages} {282} (\bibinfo {year} {2015})},\ \Eprint
  {http://arxiv.org/abs/1505.06799} {arXiv:1505.06799 [hep-ph]} \BibitemShut
  {NoStop}%
%%CITATION = ARXIV:1505.06799;%%
\bibitem [{\citenamefont {Kashiwa}\ and\ \citenamefont
  {Ohnishi}(2016)}]{Kashiwa:2016vrl}%
  \BibitemOpen
  \bibfield  {author} {\bibinfo {author} {\bibfnamefont {K.}~\bibnamefont
  {Kashiwa}}\ and\ \bibinfo {author} {\bibfnamefont {A.}~\bibnamefont
  {Ohnishi}},\ }\href {\doibase 10.1103/PhysRevD.93.116002} {\bibfield
  {journal} {\bibinfo  {journal} {Phys. Rev.}\ }\textbf {\bibinfo {volume}
  {D93}},\ \bibinfo {pages} {116002} (\bibinfo {year} {2016})},\ \Eprint
  {http://arxiv.org/abs/1602.06037} {arXiv:1602.06037 [hep-ph]} \BibitemShut
  {NoStop}%
%%CITATION = ARXIV:1602.06037;%%
\bibitem [{\citenamefont {Kashiwa}\ and\ \citenamefont
  {Ohnishi}(2017)}]{Kashiwa:2017yvy}%
  \BibitemOpen
  \bibfield  {author} {\bibinfo {author} {\bibfnamefont {K.}~\bibnamefont
  {Kashiwa}}\ and\ \bibinfo {author} {\bibfnamefont {A.}~\bibnamefont
  {Ohnishi}},\ }\href {\doibase 10.1016/j.physletb.2017.07.033} {\bibfield
  {journal} {\bibinfo  {journal} {Phys. Lett.}\ }\textbf {\bibinfo {volume}
  {B772}},\ \bibinfo {pages} {669} (\bibinfo {year} {2017})},\ \Eprint
  {http://arxiv.org/abs/1701.04953} {arXiv:1701.04953 [hep-ph]} \BibitemShut
  {NoStop}%
%%CITATION = ARXIV:1701.04953;%%
\bibitem [{\citenamefont {Wen}(1990)}]{Wen:1989iv}%
  \BibitemOpen
  \bibfield  {author} {\bibinfo {author} {\bibfnamefont {X.~G.}\ \bibnamefont
  {Wen}},\ }\href {\doibase 10.1142/S0217979290000139} {\bibfield  {journal}
  {\bibinfo  {journal} {Int.J.Mod.Phys.}\ }\textbf {\bibinfo {volume} {B4}},\
  \bibinfo {pages} {239} (\bibinfo {year} {1990})}\BibitemShut {NoStop}%
%%CITATION = IMPAE,B4,239;%%
\bibitem [{\citenamefont {Sato}(2008)}]{Sato:2007xc}%
  \BibitemOpen
  \bibfield  {author} {\bibinfo {author} {\bibfnamefont {M.}~\bibnamefont
  {Sato}},\ }\href {\doibase 10.1103/PhysRevD.77.045013} {\bibfield  {journal}
  {\bibinfo  {journal} {Phys.Rev.}\ }\textbf {\bibinfo {volume} {D77}},\
  \bibinfo {pages} {045013} (\bibinfo {year} {2008})},\ \Eprint
  {http://arxiv.org/abs/0705.2476} {arXiv:0705.2476 [hep-th]} \BibitemShut
  {NoStop}%
%%CITATION = ARXIV:0705.2476;%%
\bibitem [{\citenamefont {Roberge}\ and\ \citenamefont
  {Weiss}(1986)}]{Roberge:1986mm}%
  \BibitemOpen
  \bibfield  {author} {\bibinfo {author} {\bibfnamefont {A.}~\bibnamefont
  {Roberge}}\ and\ \bibinfo {author} {\bibfnamefont {N.}~\bibnamefont
  {Weiss}},\ }\href {\doibase 10.1016/0550-3213(86)90582-1} {\bibfield
  {journal} {\bibinfo  {journal} {Nucl.Phys.}\ }\textbf {\bibinfo {volume}
  {B275}},\ \bibinfo {pages} {734} (\bibinfo {year} {1986})}\BibitemShut
  {NoStop}%
%%CITATION = NUPHA,B275,734;%%
\bibitem [{\citenamefont {Bilgici}\ \emph {et~al.}(2008)\citenamefont
  {Bilgici}, \citenamefont {Bruckmann}, \citenamefont {Gattringer},\ and\
  \citenamefont {Hagen}}]{Bilgici:2008qy}%
  \BibitemOpen
  \bibfield  {author} {\bibinfo {author} {\bibfnamefont {E.}~\bibnamefont
  {Bilgici}}, \bibinfo {author} {\bibfnamefont {F.}~\bibnamefont {Bruckmann}},
  \bibinfo {author} {\bibfnamefont {C.}~\bibnamefont {Gattringer}}, \ and\
  \bibinfo {author} {\bibfnamefont {C.}~\bibnamefont {Hagen}},\ }\href
  {\doibase 10.1103/PhysRevD.77.094007} {\bibfield  {journal} {\bibinfo
  {journal} {Phys.Rev.}\ }\textbf {\bibinfo {volume} {D77}},\ \bibinfo {pages}
  {094007} (\bibinfo {year} {2008})},\ \Eprint {http://arxiv.org/abs/0801.4051}
  {arXiv:0801.4051 [hep-lat]} \BibitemShut {NoStop}%
%%CITATION = ARXIV:0801.4051;%%
\bibitem [{\citenamefont {de~Forcrand}\ and\ \citenamefont
  {Philipsen}(2010)}]{deForcrand:2010he}%
  \BibitemOpen
  \bibfield  {author} {\bibinfo {author} {\bibfnamefont {P.}~\bibnamefont
  {de~Forcrand}}\ and\ \bibinfo {author} {\bibfnamefont {O.}~\bibnamefont
  {Philipsen}},\ }\href {\doibase 10.1103/PhysRevLett.105.152001} {\bibfield
  {journal} {\bibinfo  {journal} {Phys. Rev. Lett.}\ }\textbf {\bibinfo
  {volume} {105}},\ \bibinfo {pages} {152001} (\bibinfo {year} {2010})},\
  \Eprint {http://arxiv.org/abs/1004.3144} {arXiv:1004.3144 [hep-lat]}
  \BibitemShut {NoStop}%
%%CITATION = ARXIV:1004.3144;%%
\bibitem [{\citenamefont {D'Elia}\ and\ \citenamefont
  {Sanfilippo}(2009)}]{D'Elia:2009qz}%
  \BibitemOpen
  \bibfield  {author} {\bibinfo {author} {\bibfnamefont {M.}~\bibnamefont
  {D'Elia}}\ and\ \bibinfo {author} {\bibfnamefont {F.}~\bibnamefont
  {Sanfilippo}},\ }\href {\doibase 10.1103/PhysRevD.80.111501} {\bibfield
  {journal} {\bibinfo  {journal} {Phys. Rev.}\ }\textbf {\bibinfo {volume}
  {D80}},\ \bibinfo {pages} {111501} (\bibinfo {year} {2009})},\ \Eprint
  {http://arxiv.org/abs/0909.0254} {arXiv:0909.0254 [hep-lat]} \BibitemShut
  {NoStop}%
%%CITATION = ARXIV:0909.0254;%%
\bibitem [{\citenamefont {Bonati}\ \emph {et~al.}(2011)\citenamefont {Bonati},
  \citenamefont {Cossu}, \citenamefont {D'Elia},\ and\ \citenamefont
  {Sanfilippo}}]{Bonati:2010gi}%
  \BibitemOpen
  \bibfield  {author} {\bibinfo {author} {\bibfnamefont {C.}~\bibnamefont
  {Bonati}}, \bibinfo {author} {\bibfnamefont {G.}~\bibnamefont {Cossu}},
  \bibinfo {author} {\bibfnamefont {M.}~\bibnamefont {D'Elia}}, \ and\ \bibinfo
  {author} {\bibfnamefont {F.}~\bibnamefont {Sanfilippo}},\ }\href {\doibase
  10.1103/PhysRevD.83.054505} {\bibfield  {journal} {\bibinfo  {journal}
  {Phys.Rev.}\ }\textbf {\bibinfo {volume} {D83}},\ \bibinfo {pages} {054505}
  (\bibinfo {year} {2011})},\ \Eprint {http://arxiv.org/abs/1011.4515}
  {arXiv:1011.4515 [hep-lat]} \BibitemShut {NoStop}%
%%CITATION = ARXIV:1011.4515;%%
\bibitem [{\citenamefont {Bonati}\ \emph {et~al.}(2016)\citenamefont {Bonati},
  \citenamefont {D'Elia}, \citenamefont {Mariti}, \citenamefont {Mesiti},
  \citenamefont {Negro},\ and\ \citenamefont {Sanfilippo}}]{Bonati:2016pwz}%
  \BibitemOpen
  \bibfield  {author} {\bibinfo {author} {\bibfnamefont {C.}~\bibnamefont
  {Bonati}}, \bibinfo {author} {\bibfnamefont {M.}~\bibnamefont {D'Elia}},
  \bibinfo {author} {\bibfnamefont {M.}~\bibnamefont {Mariti}}, \bibinfo
  {author} {\bibfnamefont {M.}~\bibnamefont {Mesiti}}, \bibinfo {author}
  {\bibfnamefont {F.}~\bibnamefont {Negro}}, \ and\ \bibinfo {author}
  {\bibfnamefont {F.}~\bibnamefont {Sanfilippo}},\ }\href {\doibase
  10.1103/PhysRevD.93.074504} {\bibfield  {journal} {\bibinfo  {journal} {Phys.
  Rev.}\ }\textbf {\bibinfo {volume} {D93}},\ \bibinfo {pages} {074504}
  (\bibinfo {year} {2016})},\ \Eprint {http://arxiv.org/abs/1602.01426}
  {arXiv:1602.01426 [hep-lat]} \BibitemShut {NoStop}%
%%CITATION = ARXIV:1602.01426;%%
\bibitem [{\citenamefont {Bonati}\ \emph {et~al.}(2019)\citenamefont {Bonati},
  \citenamefont {Calore}, \citenamefont {D'Elia}, \citenamefont {Mesiti},
  \citenamefont {Negro}, \citenamefont {Sanfilippo}, \citenamefont {Schifano},
  \citenamefont {Silvi},\ and\ \citenamefont {Tripiccione}}]{Bonati:2018fvg}%
  \BibitemOpen
  \bibfield  {author} {\bibinfo {author} {\bibfnamefont {C.}~\bibnamefont
  {Bonati}}, \bibinfo {author} {\bibfnamefont {E.}~\bibnamefont {Calore}},
  \bibinfo {author} {\bibfnamefont {M.}~\bibnamefont {D'Elia}}, \bibinfo
  {author} {\bibfnamefont {M.}~\bibnamefont {Mesiti}}, \bibinfo {author}
  {\bibfnamefont {F.}~\bibnamefont {Negro}}, \bibinfo {author} {\bibfnamefont
  {F.}~\bibnamefont {Sanfilippo}}, \bibinfo {author} {\bibfnamefont {S.~F.}\
  \bibnamefont {Schifano}}, \bibinfo {author} {\bibfnamefont {G.}~\bibnamefont
  {Silvi}}, \ and\ \bibinfo {author} {\bibfnamefont {R.}~\bibnamefont
  {Tripiccione}},\ }\href {\doibase 10.1103/PhysRevD.99.014502} {\bibfield
  {journal} {\bibinfo  {journal} {Phys. Rev.}\ }\textbf {\bibinfo {volume}
  {D99}},\ \bibinfo {pages} {014502} (\bibinfo {year} {2019})},\ \Eprint
  {http://arxiv.org/abs/1807.02106} {arXiv:1807.02106 [hep-lat]} \BibitemShut
  {NoStop}%
%%CITATION = ARXIV:1807.02106;%%
\bibitem [{\citenamefont {Goswami}\ \emph {et~al.}(2018)\citenamefont
  {Goswami}, \citenamefont {Karsch}, \citenamefont {Lahiri},\ and\
  \citenamefont {Schmidt}}]{Goswami:2018qhc}%
  \BibitemOpen
  \bibfield  {author} {\bibinfo {author} {\bibfnamefont {J.}~\bibnamefont
  {Goswami}}, \bibinfo {author} {\bibfnamefont {F.}~\bibnamefont {Karsch}},
  \bibinfo {author} {\bibfnamefont {A.}~\bibnamefont {Lahiri}}, \ and\ \bibinfo
  {author} {\bibfnamefont {C.}~\bibnamefont {Schmidt}}\ }(\bibinfo {year}
  {2018})\ \Eprint {http://arxiv.org/abs/1811.02494} {arXiv:1811.02494
  [hep-lat]} \BibitemShut {NoStop}%
%%CITATION = ARXIV:1811.02494;%%
\bibitem [{\citenamefont {Weiss}(1987)}]{Weiss:1987mp}%
  \BibitemOpen
  \bibfield  {author} {\bibinfo {author} {\bibfnamefont {N.}~\bibnamefont
  {Weiss}},\ }\href {\doibase 10.1103/PhysRevD.35.2495} {\bibfield  {journal}
  {\bibinfo  {journal} {Phys. Rev.}\ }\textbf {\bibinfo {volume} {D35}},\
  \bibinfo {pages} {2495} (\bibinfo {year} {1987})}\BibitemShut {NoStop}%
%%CITATION = PHRVA,D35,2495;%%
\bibitem [{\citenamefont {Kikuchi}()}]{kikuchi2018t}%
  \BibitemOpen
  \bibfield  {author} {\bibinfo {author} {\bibfnamefont {Y.}~\bibnamefont
  {Kikuchi}},\ }\href@noop {} {\enquote {\bibinfo {title} {{~'t {H}ooft
  anomaly, global inconsistency, and some of their applications}},}\ }\bibinfo
  {note} {Kyoto University, 2018, PhD thesis}\BibitemShut {NoStop}%
\bibitem [{\citenamefont {Yonekura}(2019)}]{Yonekura}%
  \BibitemOpen
  \bibfield  {author} {\bibinfo {author} {\bibfnamefont {K.}~\bibnamefont
  {Yonekura}},\ }\href@noop {} {\  (\bibinfo {year} {2019})},\ \Eprint
  {http://arxiv.org/abs/1901.08188} {arXiv:1901.08188 [hep-th]} \BibitemShut
  {NoStop}%
%%CITATION = ARXIV:1901.08188;%%
\bibitem [{\citenamefont {Nishimura}\ and\ \citenamefont
  {Tanizaki}(2019)}]{Nishimura:2019umw}%
  \BibitemOpen
  \bibfield  {author} {\bibinfo {author} {\bibfnamefont {H.}~\bibnamefont
  {Nishimura}}\ and\ \bibinfo {author} {\bibfnamefont {Y.}~\bibnamefont
  {Tanizaki}},\ }\href@noop {} {\  (\bibinfo {year} {2019})},\ \Eprint
  {http://arxiv.org/abs/1903.04014} {arXiv:1903.04014 [hep-th]} \BibitemShut
  {NoStop}%
%%CITATION = ARXIV:1903.04014;%%
\bibitem [{\citenamefont {Fukushima}(2004)}]{Fukushima:2003fw}%
  \BibitemOpen
  \bibfield  {author} {\bibinfo {author} {\bibfnamefont {K.}~\bibnamefont
  {Fukushima}},\ }\href {\doibase 10.1016/j.physletb.2004.04.027} {\bibfield
  {journal} {\bibinfo  {journal} {Phys.Lett.}\ }\textbf {\bibinfo {volume}
  {B591}},\ \bibinfo {pages} {277} (\bibinfo {year} {2004})},\ \Eprint
  {http://arxiv.org/abs/hep-ph/0310121} {arXiv:hep-ph/0310121 [hep-ph]}
  \BibitemShut {NoStop}%
%%CITATION = HEP-PH/0310121;%%
\bibitem [{\citenamefont {Meisinger}\ \emph {et~al.}(2002)\citenamefont
  {Meisinger}, \citenamefont {Miller},\ and\ \citenamefont
  {Ogilvie}}]{Meisinger:2001cq}%
  \BibitemOpen
  \bibfield  {author} {\bibinfo {author} {\bibfnamefont {P.~N.}\ \bibnamefont
  {Meisinger}}, \bibinfo {author} {\bibfnamefont {T.~R.}\ \bibnamefont
  {Miller}}, \ and\ \bibinfo {author} {\bibfnamefont {M.~C.}\ \bibnamefont
  {Ogilvie}},\ }\href {\doibase 10.1103/PhysRevD.65.034009} {\bibfield
  {journal} {\bibinfo  {journal} {Phys.Rev.}\ }\textbf {\bibinfo {volume}
  {D65}},\ \bibinfo {pages} {034009} (\bibinfo {year} {2002})},\ \Eprint
  {http://arxiv.org/abs/hep-ph/0108009} {arXiv:hep-ph/0108009 [hep-ph]}
  \BibitemShut {NoStop}%
%%CITATION = HEP-PH/0108009;%%
\bibitem [{\citenamefont {Dumitru}\ \emph {et~al.}(2011)\citenamefont
  {Dumitru}, \citenamefont {Guo}, \citenamefont {Hidaka}, \citenamefont
  {Altes},\ and\ \citenamefont {Pisarski}}]{Dumitru:2010mj}%
  \BibitemOpen
  \bibfield  {author} {\bibinfo {author} {\bibfnamefont {A.}~\bibnamefont
  {Dumitru}}, \bibinfo {author} {\bibfnamefont {Y.}~\bibnamefont {Guo}},
  \bibinfo {author} {\bibfnamefont {Y.}~\bibnamefont {Hidaka}}, \bibinfo
  {author} {\bibfnamefont {C.~P.~K.}\ \bibnamefont {Altes}}, \ and\ \bibinfo
  {author} {\bibfnamefont {R.~D.}\ \bibnamefont {Pisarski}},\ }\href {\doibase
  10.1103/PhysRevD.83.034022} {\bibfield  {journal} {\bibinfo  {journal}
  {Phys.Rev.}\ }\textbf {\bibinfo {volume} {D83}},\ \bibinfo {pages} {034022}
  (\bibinfo {year} {2011})},\ \Eprint {http://arxiv.org/abs/1011.3820}
  {arXiv:1011.3820 [hep-ph]} \BibitemShut {NoStop}%
%%CITATION = ARXIV:1011.3820;%%
\bibitem [{\citenamefont {Fukushima}\ and\ \citenamefont
  {Skokov}(2017)}]{Fukushima:2017csk}%
  \BibitemOpen
  \bibfield  {author} {\bibinfo {author} {\bibfnamefont {K.}~\bibnamefont
  {Fukushima}}\ and\ \bibinfo {author} {\bibfnamefont {V.}~\bibnamefont
  {Skokov}},\ }\href {\doibase 10.1016/j.ppnp.2017.05.002} {\bibfield
  {journal} {\bibinfo  {journal} {Prog. Part. Nucl. Phys.}\ }\textbf {\bibinfo
  {volume} {96}},\ \bibinfo {pages} {154} (\bibinfo {year} {2017})},\ \Eprint
  {http://arxiv.org/abs/1705.00718} {arXiv:1705.00718 [hep-ph]} \BibitemShut
  {NoStop}%
%%CITATION = ARXIV:1705.00718;%%
\bibitem [{\citenamefont {Kashiwa}\ \emph {et~al.}(2008)\citenamefont
  {Kashiwa}, \citenamefont {Kouno}, \citenamefont {Matsuzaki},\ and\
  \citenamefont {Yahiro}}]{Kashiwa:2007hw}%
  \BibitemOpen
  \bibfield  {author} {\bibinfo {author} {\bibfnamefont {K.}~\bibnamefont
  {Kashiwa}}, \bibinfo {author} {\bibfnamefont {H.}~\bibnamefont {Kouno}},
  \bibinfo {author} {\bibfnamefont {M.}~\bibnamefont {Matsuzaki}}, \ and\
  \bibinfo {author} {\bibfnamefont {M.}~\bibnamefont {Yahiro}},\ }\href
  {\doibase 10.1016/j.physletb.2008.01.075} {\bibfield  {journal} {\bibinfo
  {journal} {Phys.Lett.}\ }\textbf {\bibinfo {volume} {B662}},\ \bibinfo
  {pages} {26} (\bibinfo {year} {2008})},\ \Eprint
  {http://arxiv.org/abs/0710.2180} {arXiv:0710.2180 [hep-ph]} \BibitemShut
  {NoStop}%
%%CITATION = ARXIV:0710.2180;%%
\bibitem [{\citenamefont {Roessner}\ \emph {et~al.}(2007)\citenamefont
  {Roessner}, \citenamefont {Ratti},\ and\ \citenamefont
  {Weise}}]{Roessner:2006xn}%
  \BibitemOpen
  \bibfield  {author} {\bibinfo {author} {\bibfnamefont {S.}~\bibnamefont
  {Roessner}}, \bibinfo {author} {\bibfnamefont {C.}~\bibnamefont {Ratti}}, \
  and\ \bibinfo {author} {\bibfnamefont {W.}~\bibnamefont {Weise}},\ }\href
  {\doibase 10.1103/PhysRevD.75.034007} {\bibfield  {journal} {\bibinfo
  {journal} {Phys.Rev.}\ }\textbf {\bibinfo {volume} {D75}},\ \bibinfo {pages}
  {034007} (\bibinfo {year} {2007})},\ \Eprint
  {http://arxiv.org/abs/hep-ph/0609281} {arXiv:hep-ph/0609281 [hep-ph]}
  \BibitemShut {NoStop}%
%%CITATION = HEP-PH/0609281;%%
\bibitem [{\citenamefont {Sakai}\ \emph {et~al.}(2008)\citenamefont {Sakai},
  \citenamefont {Kashiwa}, \citenamefont {Kouno},\ and\ \citenamefont
  {Yahiro}}]{Sakai:2008py}%
  \BibitemOpen
  \bibfield  {author} {\bibinfo {author} {\bibfnamefont {Y.}~\bibnamefont
  {Sakai}}, \bibinfo {author} {\bibfnamefont {K.}~\bibnamefont {Kashiwa}},
  \bibinfo {author} {\bibfnamefont {H.}~\bibnamefont {Kouno}}, \ and\ \bibinfo
  {author} {\bibfnamefont {M.}~\bibnamefont {Yahiro}},\ }\href {\doibase
  10.1103/PhysRevD.77.051901} {\bibfield  {journal} {\bibinfo  {journal}
  {Phys.Rev.}\ }\textbf {\bibinfo {volume} {D77}},\ \bibinfo {pages} {051901}
  (\bibinfo {year} {2008})},\ \Eprint {http://arxiv.org/abs/0801.0034}
  {arXiv:0801.0034 [hep-ph]} \BibitemShut {NoStop}%
%%CITATION = ARXIV:0801.0034;%%
\bibitem [{\citenamefont {Kashiwa}\ \emph {et~al.}(2013)\citenamefont
  {Kashiwa}, \citenamefont {Sasaki}, \citenamefont {Kouno},\ and\ \citenamefont
  {Yahiro}}]{Kashiwa:2012xm}%
  \BibitemOpen
  \bibfield  {author} {\bibinfo {author} {\bibfnamefont {K.}~\bibnamefont
  {Kashiwa}}, \bibinfo {author} {\bibfnamefont {T.}~\bibnamefont {Sasaki}},
  \bibinfo {author} {\bibfnamefont {H.}~\bibnamefont {Kouno}}, \ and\ \bibinfo
  {author} {\bibfnamefont {M.}~\bibnamefont {Yahiro}},\ }\href {\doibase
  10.1103/PhysRevD.87.016015} {\bibfield  {journal} {\bibinfo  {journal} {Phys.
  Rev.}\ }\textbf {\bibinfo {volume} {D87}},\ \bibinfo {pages} {016015}
  (\bibinfo {year} {2013})},\ \Eprint {http://arxiv.org/abs/1208.2283}
  {arXiv:1208.2283 [hep-ph]} \BibitemShut {NoStop}%
%%CITATION = ARXIV:1208.2283;%%
\bibitem [{\citenamefont {Shimizu}\ and\ \citenamefont
  {Yonekura}(2018)}]{Shimizu:2017asf}%
  \BibitemOpen
  \bibfield  {author} {\bibinfo {author} {\bibfnamefont {H.}~\bibnamefont
  {Shimizu}}\ and\ \bibinfo {author} {\bibfnamefont {K.}~\bibnamefont
  {Yonekura}},\ }\href {\doibase 10.1103/PhysRevD.97.105011} {\bibfield
  {journal} {\bibinfo  {journal} {Phys. Rev.}\ }\textbf {\bibinfo {volume}
  {D97}},\ \bibinfo {pages} {105011} (\bibinfo {year} {2018})},\ \Eprint
  {http://arxiv.org/abs/1706.06104} {arXiv:1706.06104 [hep-th]} \BibitemShut
  {NoStop}%
%%CITATION = ARXIV:1706.06104;%%
\bibitem [{\citenamefont {Ishiyama}\ \emph {et~al.}(2010)\citenamefont
  {Ishiyama}, \citenamefont {Murata}, \citenamefont {So},\ and\ \citenamefont
  {Takenaga}}]{Ishiyama:2009bk}%
  \BibitemOpen
  \bibfield  {author} {\bibinfo {author} {\bibfnamefont {K.}~\bibnamefont
  {Ishiyama}}, \bibinfo {author} {\bibfnamefont {M.}~\bibnamefont {Murata}},
  \bibinfo {author} {\bibfnamefont {H.}~\bibnamefont {So}}, \ and\ \bibinfo
  {author} {\bibfnamefont {K.}~\bibnamefont {Takenaga}},\ }\href {\doibase
  10.1143/PTP.123.257} {\bibfield  {journal} {\bibinfo  {journal} {Prog. Theor.
  Phys.}\ }\textbf {\bibinfo {volume} {123}},\ \bibinfo {pages} {257} (\bibinfo
  {year} {2010})},\ \Eprint {http://arxiv.org/abs/0911.4555} {arXiv:0911.4555
  [hep-lat]} \BibitemShut {NoStop}%
%%CITATION = ARXIV:0911.4555;%%
\bibitem [{\citenamefont {Kashiwa}\ \emph {et~al.}(2009)\citenamefont
  {Kashiwa}, \citenamefont {Kouno},\ and\ \citenamefont
  {Yahiro}}]{Kashiwa:2009ki}%
  \BibitemOpen
  \bibfield  {author} {\bibinfo {author} {\bibfnamefont {K.}~\bibnamefont
  {Kashiwa}}, \bibinfo {author} {\bibfnamefont {H.}~\bibnamefont {Kouno}}, \
  and\ \bibinfo {author} {\bibfnamefont {M.}~\bibnamefont {Yahiro}},\ }\href
  {\doibase 10.1103/PhysRevD.80.117901} {\bibfield  {journal} {\bibinfo
  {journal} {Phys.Rev.}\ }\textbf {\bibinfo {volume} {D80}},\ \bibinfo {pages}
  {117901} (\bibinfo {year} {2009})},\ \Eprint {http://arxiv.org/abs/0908.1213}
  {arXiv:0908.1213 [hep-ph]} \BibitemShut {NoStop}%
%%CITATION = ARXIV:0908.1213;%%
\bibitem [{\citenamefont {Bilgici}\ \emph {et~al.}(2010)\citenamefont
  {Bilgici}, \citenamefont {Bruckmann}, \citenamefont {Danzer}, \citenamefont
  {Gattringer}, \citenamefont {Hagen}, \citenamefont {Ilgenfritz},\ and\
  \citenamefont {Maas}}]{Bilgici:2009tx}%
  \BibitemOpen
  \bibfield  {author} {\bibinfo {author} {\bibfnamefont {E.}~\bibnamefont
  {Bilgici}}, \bibinfo {author} {\bibfnamefont {F.}~\bibnamefont {Bruckmann}},
  \bibinfo {author} {\bibfnamefont {J.}~\bibnamefont {Danzer}}, \bibinfo
  {author} {\bibfnamefont {C.}~\bibnamefont {Gattringer}}, \bibinfo {author}
  {\bibfnamefont {C.}~\bibnamefont {Hagen}}, \bibinfo {author} {\bibfnamefont
  {E.~M.}\ \bibnamefont {Ilgenfritz}}, \ and\ \bibinfo {author} {\bibfnamefont
  {A.}~\bibnamefont {Maas}},\ }\href {\doibase 10.1007/s00601-009-0068-x}
  {\bibfield  {journal} {\bibinfo  {journal} {Few Body Syst.}\ }\textbf
  {\bibinfo {volume} {47}},\ \bibinfo {pages} {125} (\bibinfo {year} {2010})},\
  \Eprint {http://arxiv.org/abs/0906.3957} {arXiv:0906.3957 [hep-lat]}
  \BibitemShut {NoStop}%
%%CITATION = ARXIV:0906.3957;%%
\bibitem [{\citenamefont {Bilgici}()}]{Bilgici:2009phd}%
  \BibitemOpen
  \bibfield  {author} {\bibinfo {author} {\bibfnamefont {E.}~\bibnamefont
  {Bilgici}},\ }\href@noop {} {\enquote {\bibinfo {title} {{Signatures of
  confinement and chiral symmetry breaking in spectral quantities of lattice
  Dirac operators}},}\ }\bibinfo {note} {University of Graz, 2009,
  (http://physik.uni-graz.at/itp/files/bilgici/dissertation.pdf)}\BibitemShut
  {NoStop}%
\bibitem [{\citenamefont {Bruckmann}\ and\ \citenamefont
  {Endrodi}(2011)}]{Bruckmann:2011zx}%
  \BibitemOpen
  \bibfield  {author} {\bibinfo {author} {\bibfnamefont {F.}~\bibnamefont
  {Bruckmann}}\ and\ \bibinfo {author} {\bibfnamefont {G.}~\bibnamefont
  {Endrodi}},\ }\href {\doibase 10.1103/PhysRevD.84.074506} {\bibfield
  {journal} {\bibinfo  {journal} {Phys. Rev.}\ }\textbf {\bibinfo {volume}
  {D84}},\ \bibinfo {pages} {074506} (\bibinfo {year} {2011})},\ \Eprint
  {http://arxiv.org/abs/1104.5664} {arXiv:1104.5664 [hep-lat]} \BibitemShut
  {NoStop}%
%%CITATION = ARXIV:1104.5664;%%
\bibitem [{\citenamefont {Fischer}(2009)}]{Fischer:2009wc}%
  \BibitemOpen
  \bibfield  {author} {\bibinfo {author} {\bibfnamefont {C.~S.}\ \bibnamefont
  {Fischer}},\ }\href {\doibase 10.1103/PhysRevLett.103.052003} {\bibfield
  {journal} {\bibinfo  {journal} {Phys.Rev.Lett.}\ }\textbf {\bibinfo {volume}
  {103}},\ \bibinfo {pages} {052003} (\bibinfo {year} {2009})},\ \Eprint
  {http://arxiv.org/abs/0904.2700} {arXiv:0904.2700 [hep-ph]} \BibitemShut
  {NoStop}%
%%CITATION = ARXIV:0904.2700;%%
\bibitem [{\citenamefont {Fischer}\ and\ \citenamefont
  {Mueller}(2009)}]{Fischer:2009gk}%
  \BibitemOpen
  \bibfield  {author} {\bibinfo {author} {\bibfnamefont {C.~S.}\ \bibnamefont
  {Fischer}}\ and\ \bibinfo {author} {\bibfnamefont {J.~A.}\ \bibnamefont
  {Mueller}},\ }\href {\doibase 10.1103/PhysRevD.80.074029} {\bibfield
  {journal} {\bibinfo  {journal} {Phys. Rev.}\ }\textbf {\bibinfo {volume}
  {D80}},\ \bibinfo {pages} {074029} (\bibinfo {year} {2009})},\ \Eprint
  {http://arxiv.org/abs/0908.0007} {arXiv:0908.0007 [hep-ph]} \BibitemShut
  {NoStop}%
%%CITATION = ARXIV:0908.0007;%%
\bibitem [{\citenamefont {Gatto}\ and\ \citenamefont
  {Ruggieri}(2010)}]{Gatto:2010qs}%
  \BibitemOpen
  \bibfield  {author} {\bibinfo {author} {\bibfnamefont {R.}~\bibnamefont
  {Gatto}}\ and\ \bibinfo {author} {\bibfnamefont {M.}~\bibnamefont
  {Ruggieri}},\ }\href {\doibase 10.1103/PhysRevD.82.054027} {\bibfield
  {journal} {\bibinfo  {journal} {Phys. Rev.}\ }\textbf {\bibinfo {volume}
  {D82}},\ \bibinfo {pages} {054027} (\bibinfo {year} {2010})},\ \Eprint
  {http://arxiv.org/abs/1007.0790} {arXiv:1007.0790 [hep-ph]} \BibitemShut
  {NoStop}%
%%CITATION = ARXIV:1007.0790;%%
\bibitem [{\citenamefont {Zhang}\ and\ \citenamefont
  {Miao}(2015)}]{Zhang:2015baa}%
  \BibitemOpen
  \bibfield  {author} {\bibinfo {author} {\bibfnamefont {Z.}~\bibnamefont
  {Zhang}}\ and\ \bibinfo {author} {\bibfnamefont {Q.}~\bibnamefont {Miao}},\
  }\href@noop {} {\  (\bibinfo {year} {2015})},\ \Eprint
  {http://arxiv.org/abs/1507.07224} {arXiv:1507.07224 [hep-ph]} \BibitemShut
  {NoStop}%
%%CITATION = ARXIV:1507.07224;%%
\bibitem [{\citenamefont {Zhang}\ and\ \citenamefont
  {Lu}(2017)}]{Zhang:2017bem}%
  \BibitemOpen
  \bibfield  {author} {\bibinfo {author} {\bibfnamefont {Z.}~\bibnamefont
  {Zhang}}\ and\ \bibinfo {author} {\bibfnamefont {H.}~\bibnamefont {Lu}},\
  }\href@noop {} {\  (\bibinfo {year} {2017})},\ \Eprint
  {http://arxiv.org/abs/1705.09953} {arXiv:1705.09953 [hep-ph]} \BibitemShut
  {NoStop}%
%%CITATION = ARXIV:1705.09953;%%
\end{thebibliography}%

\end{document}